\RequirePackage{rotating}
\documentclass[a4paper,fleqn,usenatbib]{mnras}
\usepackage{graphicx}
\usepackage{longtable}
\usepackage{rotate}
\usepackage{rotating}
\usepackage{mathtools}
\usepackage{pdflscape}
\usepackage{scalerel}
\usepackage{times}
\usepackage{verbatim}
\newif\ifAMStwofonts

\def\gs{\mathrel{\hbox{\rlap{\hbox{\lower4pt\hbox{$\sim$}}}\hbox{$>$}}}}
\def\ls{\mathrel{\hbox{\rlap{\hbox{\lower4pt\hbox{$\sim$}}}\hbox{$<$}}}}

\def\suzaku{{\it Suzaku}}
\def\nustar{{\it NuSTAR}}

\def\swift{{\it Swift}}
\def\xmm{{\it XMM-Newton}}

\def\et{{et al.\ }}

\def\wkk{{WKK~4438}}
\def\iras13{{IRAS~13224--3809}}
\def\1h07{{1H~0707--495}}
\def\izw1{{I~Zw~1}}
\def\rg{{\thinspace r_{\rm g}}}
\def\risco{{\thinspace r_{\rm ISCO}}}

\def\Cdof{{C/{\rm dof}}}
\def\delC{{\Delta{\rm C}}}
\def\delDIC{{\Delta{\rm DIC}}}

\def\feka{{Fe~K$\alpha$}}

\def\feii{{Fe~\textsc{ii}}}

\def\nh{{N_{\rm H}}}

\def\deg{^{\circ}}

\def\cm{{\rm\thinspace cm}}
\def\erg{{\rm\thinspace erg}}
\def\eV{{\rm\thinspace eV}}

\def\keV{{\rm\thinspace keV}}

\def\Msun{\hbox{$\rm\thinspace M_{\odot}$}}

\def\s{{\rm\thinspace s}}
\def\ks{{\rm\thinspace ks}}

\def\cmps{\hbox{$\cm\s^{-1}\,$}}

\def\pscm{\hbox{$\cm^{-2}\,$}}

\def\pccm{\hbox{$\cm^{-3}\,$}}

\title[A truncated inner disc in the Seyfert 1  \wkk\ ]
      {
A truncated inner disc in the Seyfert 1 galaxy \wkk\    }

\author[L. C. Gallo \et]
       {
       L. C. Gallo,$^1$ 
       M. Z. Buhariwalla,$^1$
       J. Jiang,$^2$
       F. D'Ammando$^3$ and
       D. J. Walton$^{4,2}$  
        \\ 
$^{1}$ Department of Astronomy and Physics, Saint Mary's University, 923 Robie Street, Halifax, NS, B3H 3C3, Canada \\
$^{2}$ Institute of Astronomy, University of Cambridge, Madingley Road, Cambridge CB3 0HA\\
$^{3}$ INAF - Istituto di Radioastronomia, Via Gobetti 101, I-40129 Bologna, Italy \\
$^{4}$ Centre for Astrophysics Research, University of Hertfordshire, College Lane, Hatfield AL10 9AB, UK  \\
}
\date{Accepted. Received. }
\pagerange{\pageref{firstpage}--\pageref{lastpage}}
\pubyear{2022}
\begin{document}
\maketitle
\label{firstpage}

\begin{abstract}
Understanding if and when the accretion disc extends down to the innermost stable circular orbit (ISCO) is important since it is the fundamental assumption behind measuring black hole spin.  Here, we examine the 2013 and 2018 \nustar\ and \swift\ data ($0.5-50\keV$) of the narrow-line Seyfert 1 galaxy, \wkk.  The X-ray emission can be fitted well with models depicting a corona and blurred reflection originating from a disc around a low spin ($a_*\approx0$) black hole.  However, such models result in unconventional values for some of the parameters (e.g. inverted emissivity profile and high coronal height).  Alternatively, equally good fits can be achieved if the disc is truncated at $\sim10\rg$ and the black hole is spinning at the Thorne limit ($a_*=0.998$).  In these cases, the model parameters are consistent with the interpretation that the corona is centrally located close to the black hole and illuminating the disc at a larger distance.   

\end{abstract}



\section{Introduction}
\label{sect:intro}
Disc truncation is a common explanation for describing the low-luminosity phases of the hard state of stellar mass black hole binaries and the behaviour of low-luminosity active galactic nuclei (LLAGN).  In the case of LLAGN, the standard disc is truncated at large radii and the inner region is described by a hot, slow accretion flow that is radiatively inefficient (e.g. \citealt{ADAF, Lasota, Gammie99, Ptak04}).  

Disc truncation is less common and perhaps unexpected in objects that are radiating efficiently and where the flow can be described by an optically-thick, geometrically-thin $\alpha$-disc \citep{Shakura73}.  Indeed, these are exactly the objects where the black hole spin parameter ($a_*=a/M=Jc/GM^2$) can be measured well using thermal continuum fitting and \feka\ modelling (e.g. \citealt{Shafee06, Brenneman06}; see \citealt{Reynolds21} for a recent review).  Both of these techniques rely on the assumption that the disc edge extends down to the innermost stable circular orbit (ISCO).  Therefore, it is important to examine if there are radiatively efficient AGN where this assumption does not hold.

\wkk\ (IGR~J14552--5133, LEDA~$3076910$; $z=0.016$, \citealt{Masetti06}) is located toward the Galactic centre and viewed through a Galactic column density of $4.34\times10^{21}\pscm$ \citep{Willingale}. The  AGN is relatively bright and it was detected in the \swift-BAT survey \citep{Oh18}.  Based on its optical emission line properties \citep{Masetti06}, it is formally defined as a narrow-line Seyfert~1 (NLS1), however, its X-ray properties are less extreme than typical NLS1s (e.g. \citealt{Waddell20, Waddell22, Grupe04a, Grupe04b}).  The most detailed X-ray study of the object was based on the 2012 \suzaku\ and short 2013 \nustar\ observation by \cite{JiangWKK}.  They found tentative evidence of an ultrafast outflow and indication for a large inner disc radius.

In this work, the $0.5-50\keV$ emission of \wkk\ is examined using new \nustar\ and \swift\ data from 2018 in conjunction with the previous 2013 data.  The data are described in the following section.  The variability and spectral analysis are presented in Sections~\ref{sect:lcurve} and \ref{sect:spec}, respectively.  Discussion and concluding remarks are in Section~\ref{sect:dis}.


\section{Observation and data reduction}
\label{sect:data}

\wkk\ was observed with \nustar\ \citep{nustar} on two occasions.  The first time on 19 September 2013 for a duration of $\sim40\ks$ and the second time starting on 21 September 2018  for a duration of  $\sim200\ks$ (PI: Jiang, J.).  At both epochs, a snap-shot observation was obtained with the X-ray Telescope (XRT, \citealt{xrt}) on the {\it Neil Gehrels} \swift\ observatory \citep{swift} . 
 A summary of the observations used in this work is provided in Table~\ref{tab:obslog}.
            \begin{table}
            \begin{center}
            \caption{The observation log of the X-ray spectral data utilised in this analysis of \wkk.
            The observatories and instruments used are listed in column (1).  The ID corresponding to the observation is given in column (2).  The start date and total good-time exposure follow in columns (3) and (4).  The total source counts (background corrected) are given in column (5) for the combined FBMA+B  in 2018 ($4-50\keV$) and 2013 ($4-30\keV$), and for the XRT ($0.5-4\keV$). 
            }
            \begin{tabular}{ccccc}                
            \hline
            (1) & (2) & (3) & (4) & (5)   \\
            Observatory & Observation  & Start Date  &   Exposure  & Source  \\
             (Instruments) & ID & (year.mm.dd)   &   (ks)    & Counts     \\
			\hline
            \hline
            \nustar\ & 60061259002 & 2013.09.19 &  21.9  &  5985 \\
      (FPMA+B)   & 60401022002 & 2018.09.21 &  100.9  &  33515 \\
      \\
            \swift\   & 00080140001   & 2013.09.20 &  7.1  &  639  \\
         (XRT) & 00088730001   & 2018.09.22 &  2.1  &  214  \\
            \hline
            \label{tab:obslog}
            \end{tabular}
            \end{center}
            \end{table}

The \swift~XRT spectra ($0.3-10\keV$) were generated using the XRT data products generator\footnote{https://www.swift.ac.uk/user\_objects}   \citep{Evans09} for the specific dates of the \nustar\ observations.  The XRT was operated in photon counting mode.  The \nustar\ data were processed in the standard manner\footnote{https://heasarc.gsfc.nasa.gov/docs/nustar/analysis/nustar\_swguide.pdf} with the \nustar\ Data Analysis Software {\sc nustardas v2.0.0} and calibration files from the \nustar\ {\sc caldb v20200811}. Both focal plane modules (FPMA and FPMA) operated normally and generated data.  Unfiltered event files were cleaned with {\sc nupipeline} with the default settings on mode 1 data only.  Light curves and spectra were extracted from the data using {\sc nuproducts}, which also creates the spectral response matrices.  \nustar\ source data were extracted from a circular region with a radius of $75\arcsec$ centred on the source. The background was selected from a nearby circular region with a radius of $110\arcsec$.  

The 2012 \suzaku\ light curve ($0.5-10\keV$) of \wkk\ from the combined front-illuminated CCDs is presented in Fig.~\ref{fig:lcurve}.  The data come from products generated by \cite{Waddell20} and are presented here for comparison.  Detailed examination of the \suzaku\ spectra was carried out by \cite{JiangWKK}.

\section{Rapid flux variations}
\label{sect:lcurve}
\begin{figure*}
   \centering
   {\scalebox{0.49}{\includegraphics[trim= 0cm 0cm 1cm 2.cm, angle=0,clip=true]{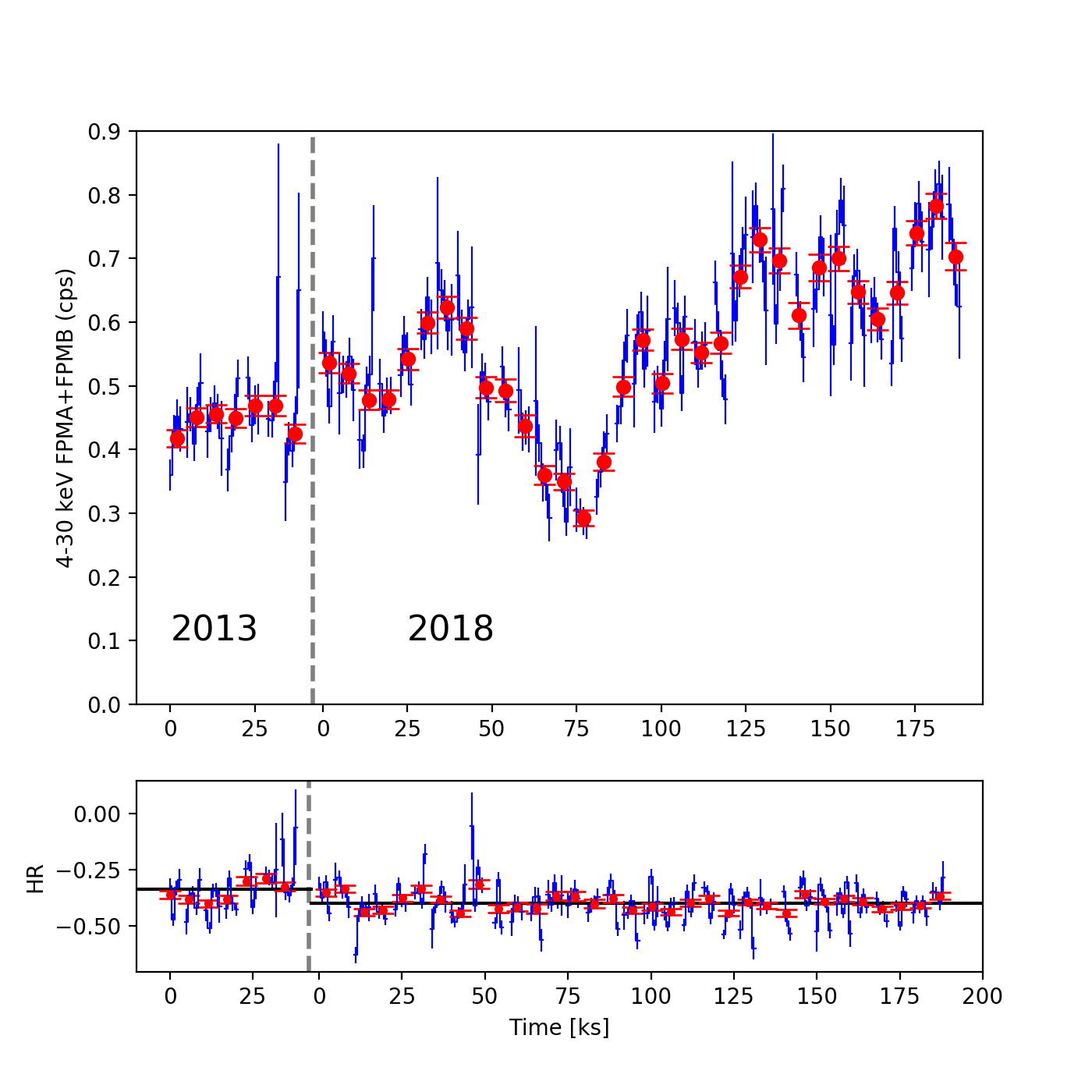}}}
   {\scalebox{0.49}{\includegraphics[trim= 0cm 0cm 1cm 2cm, clip=true]{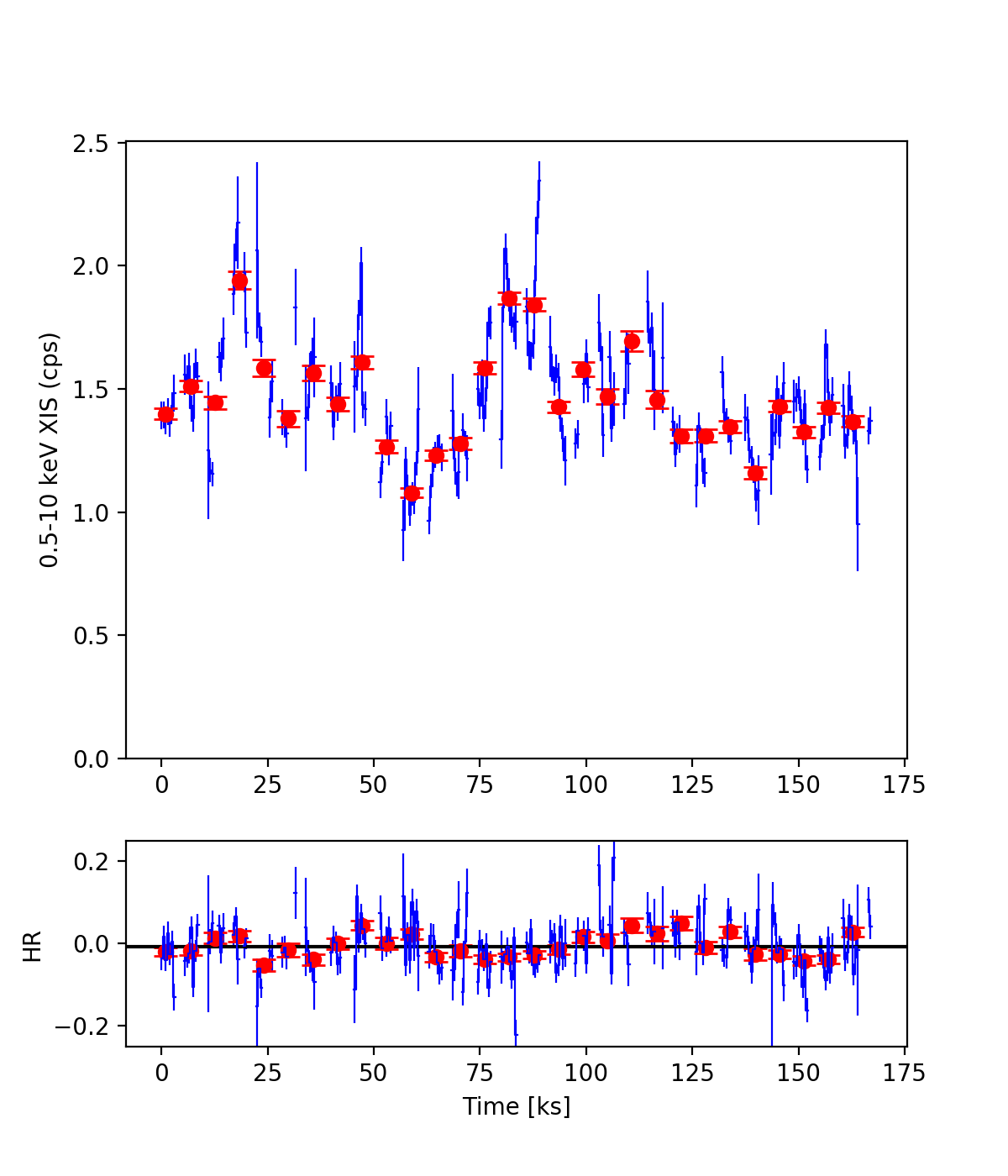}}}
   \caption{ \nustar\ (left) and \suzaku\ (right) light curves (upper panels) and hardness ratio curves where $HR=(H-S)/(H+S)$ (lower panels).  The blue points are $1000\s$ bins and the red points are orbital bins ($5760\s$). For the \nustar\ hardness ratios, $H=10-30\keV$ and $S=4-10\keV$. For \suzaku, $H=2-10\keV$ and $S=0.5-2\keV$.}
   \label{fig:lcurve}
\end{figure*}

The 2013 and 2018 \nustar\  ($4-30\keV$) light curves  as well as the 2012 \suzaku\ ($0.5-10\keV$) light curves from the combined front-illuminated CCDs are shown in Fig.~\ref{fig:lcurve}.  Substantial variations, up to $\pm50$ per cent, are present in both energy ranges and on all time scales probed with the data (i.e. $\ks$ to days).  However, spectral variations as demonstrated from hardness ratios (Fig.~\ref{fig:lcurve}) and fractional variability analysis are modest.  The \nustar\ spectra between 2013 and 2018 are quite similar.  A difference spectrum between the two epochs can be fitted between $4-30\keV$ with a single power law ($\Gamma\sim2.1$) indicating the differences are dominated by flux changes (Fig.~\ref{fig:spec}).

For a black hole mass of $2\times10^6 \Msun$ in \wkk\ \citep{Malizia08}, $10\rg$ corresponds to a light-crossing time of $\sim100\s$ and a dynamical time scale of $\sim 2\ks$.  The high-amplitude and rapid variations on a kilosecond time scale in \wkk\ indicate that the X-rays are originating from a compact region.  
The insignificant spectral variations would suggest that only one component might dominate the X-ray band or that multiple components vary in such a way as to not change the overall shape of the spectrum.

\section{Spectral analysis}
\label{sect:spec}

\begin{figure*}
\begin{center}
\begin{minipage}{0.45\linewidth}
\scalebox{0.33}{\includegraphics[trim= 1cm 1.5cm 2cm 3cm, angle=0,clip=true]{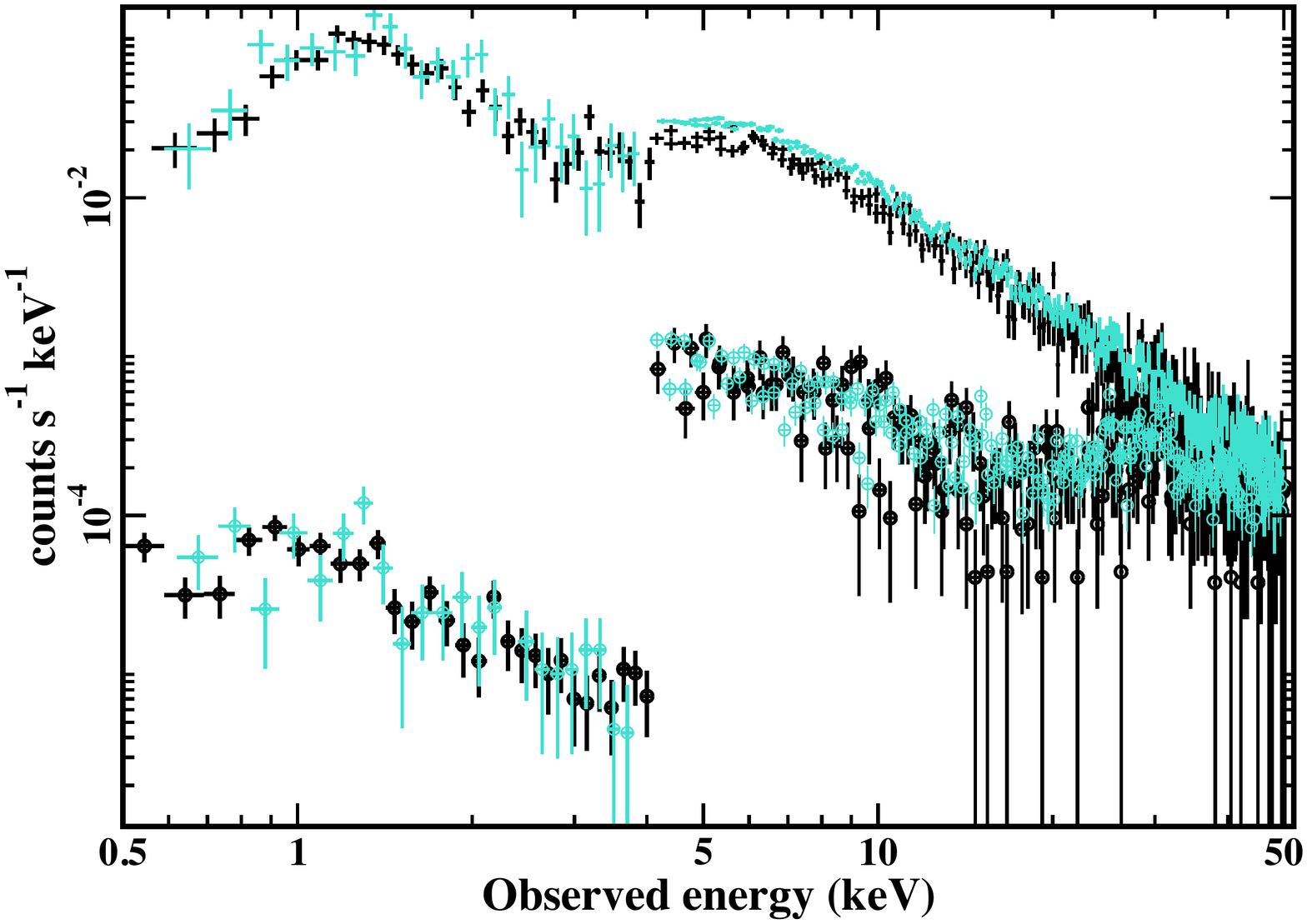}}
\scalebox{0.33}{\includegraphics[trim= 1cm 1.5cm 2cm 5cm, clip=true,angle=0]{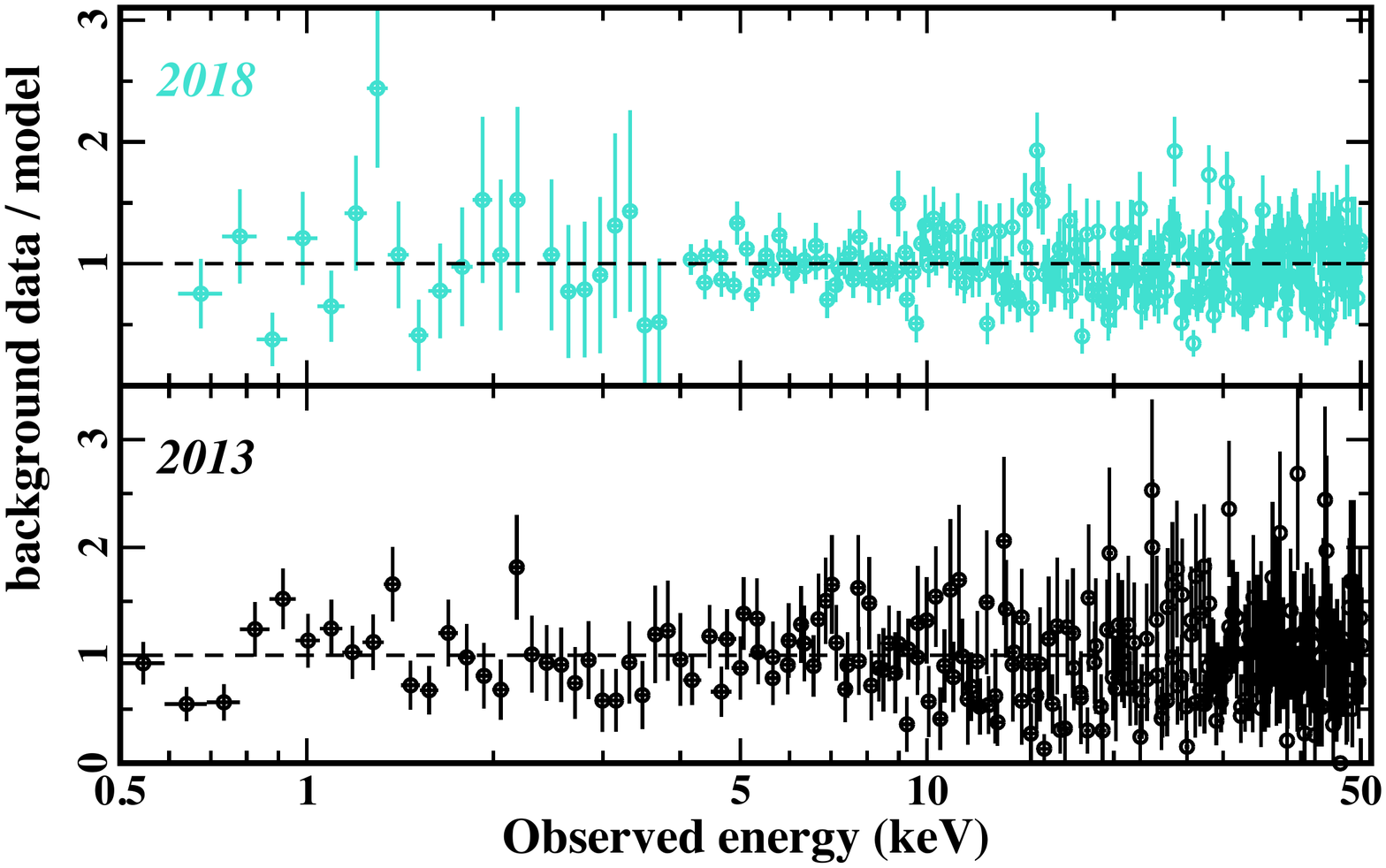}}
\end{minipage}  \hfill
\begin{minipage}{0.45\linewidth}
\scalebox{0.33}{\includegraphics[trim= 1cm 1.5cm 2cm 3cm, angle=0,clip=true]{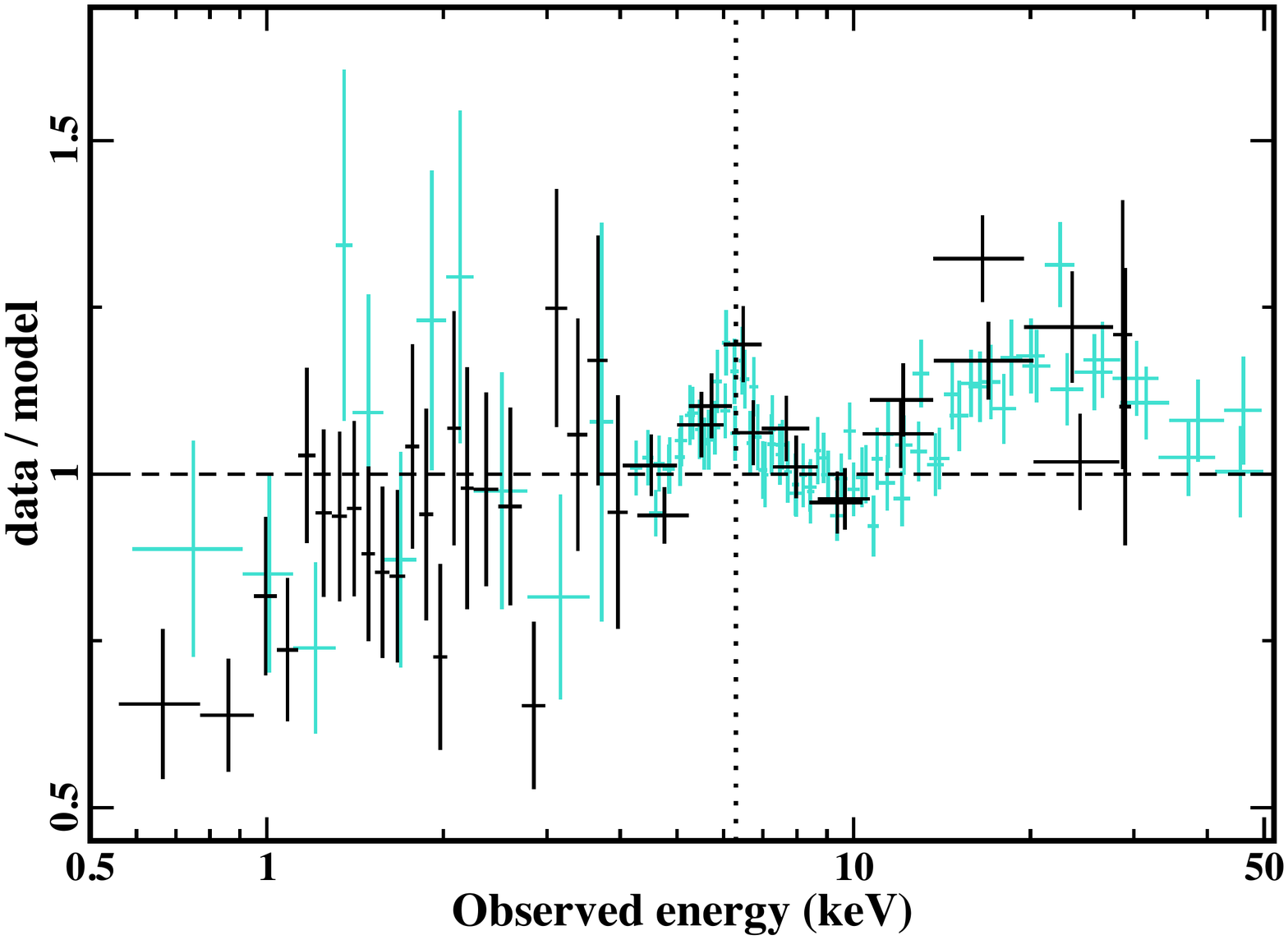}}
\scalebox{0.33}{\includegraphics[trim= 1cm 1.5cm 5cm 1cm, clip=true,angle=90]{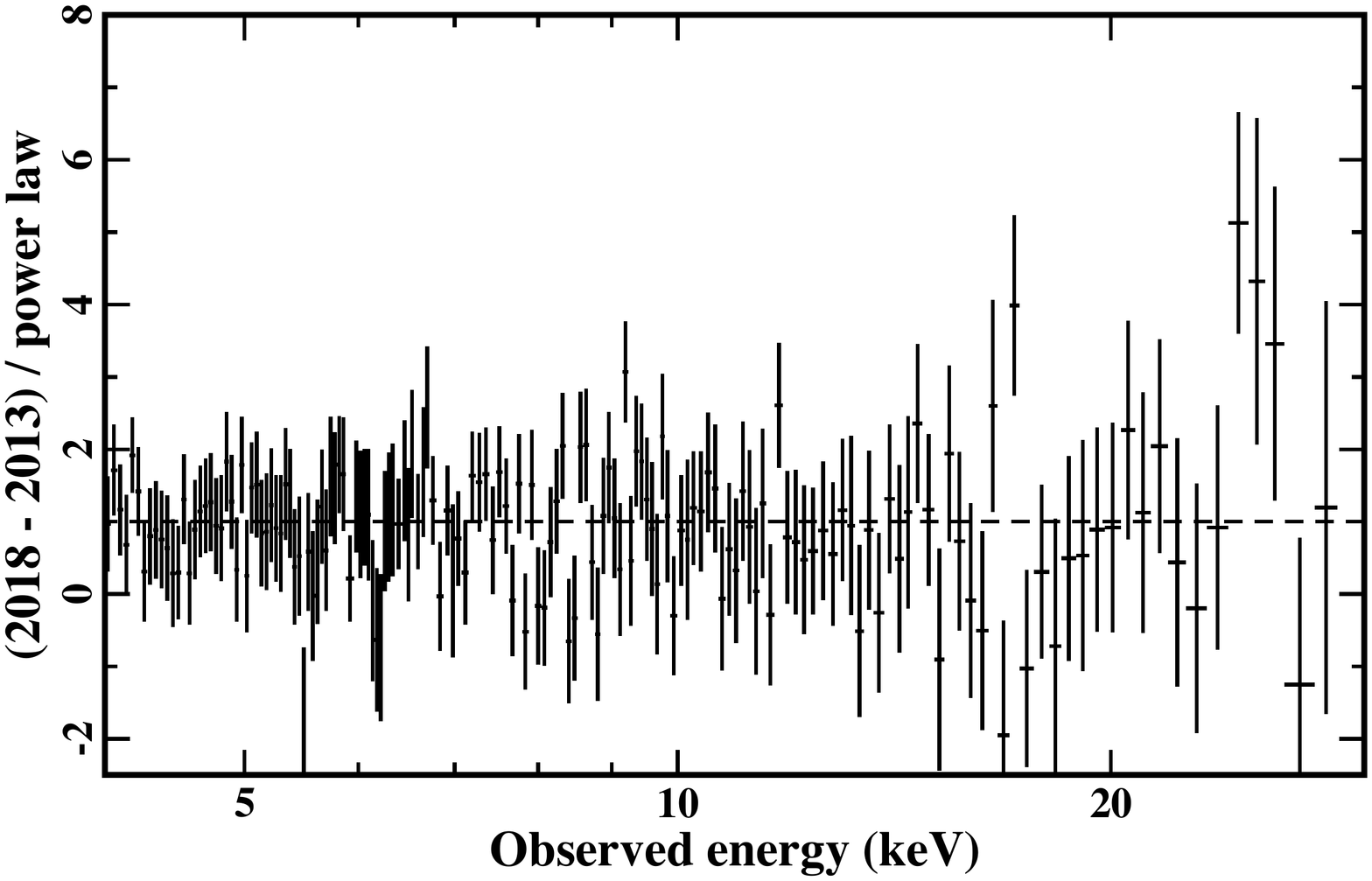}}
\end{minipage}
\end{center}
\caption{The spectral data from the 2013 (black) and 2018 (turquoise) observations of \wkk.  The \nustar\  data are above $4\keV$ and the \swift~XRT data are below $4\keV$.  The corresponding background spectra are shown with circular data points. {\bf Upper left:} The spectral data from each epoch and telescope compared to respective background levels.  Data from FPMA and FPMB are plotted, but shown as the same colour to ease comparison between epochs.   {\bf Lower left:}  The data-to-model ratio for the background spectra of each instrument.  The background model is phenomenological comprising of a blackbody, power law, and Gaussian profiles for instrumental emission lines.  {\bf Upper right:} The data-to-model ratio of an absorbed power law fitted to the spectra between $4-5$, $7.5-10$, and $40-50\keV$, and extrapolated over the entire band.  Excess emission is evident between $10-40\keV$, and in the \feka\ band.  The vertical dotted-line marks $6.4\keV$ in the source frame.  The spectrum below $\sim1.5\keV$ is lower than predicted from the extrapolated power law. {\bf Lower right:}  The difference spectrum between the 2018 and 2013 \nustar\ data can be fitted with a single power law ($\Gamma\sim2.1$). The data from each instrument is combined at each epoch.
}
\label{fig:spec}
\end{figure*}

All spectra are grouped using optimal binning \citep{optbin}.   To utilise Cash statistics (\citealt{Cash}, see below) effectively, the background spectra are fitted separately with a  phenomenological model comprising of a blackbody, power law, and Gaussian profiles for instrumental emission lines (Fig.~\ref{fig:spec}).   The data are analysed in regions where the source is dominant over the background and the calibration is well understood.  For the 2013 and 2018 \nustar\ data this corresponds to $4-30\keV$ and $4-50\keV$, respectively.  To constrain the low-energy emission, \swift~XRT data obtained simultaneously with each \nustar\ observation are used between $0.5-4\keV$.  Given the low count rate, many XRT bins above $4\keV$ have zero counts.  The parameters are linked between data sets at each epoch except for a calibration constant between the two FPM detectors and another constant between \nustar\ and \swift.  Both calibration constants are in agreement with the expected values.  Both epochs are fitted simultaneously.

Spectral fitting was performed using {\sc xspec v12.11.1} and the fit quality was evaluated by minimising the modified C-statistic (based on the Cash statistic, \citealt{Cash, Humphrey09}) in {\sc xspec}.  Parameter uncertainties for the best-fit models were determined using a Markov Chain Monte Carlo (MCMC).  Using the {\sc xspec} implementation of the \cite{mcmc} algorithm, the MCMC was run with 100 walkers for $2.2\times10^5$ steps and burning the first 20000 steps.  The reported values are the mean of the likelihood function and the uncertainties correspond to the 90 per cent confidence interval.  

For the final models, the Deviance Information Criterion (DIC; \citealt{dic}) is used to compare different models and quantify a goodness-of-fit (the C-statistic is used for preliminary models in limited energy ranges).  Using the MCMC posterior distribution, the DIC adds a penalty for the effective number of free parameters (i.e. half the variance of the distribution, \citealt{dic2}) to the posterior mean fit statistic.  A smaller DIC value favours one model over another.  A difference in DIC ($\delDIC$) between 0 and 2 indicates a marginal preference for a model whereas $\delDIC>6$ reflects a strong preference \citep{dic3}.

All parameters are reported in the rest frame of the source unless specified otherwise, but figures remain in the observed frame.  A value of the total Galactic column density, considering atomic and molecular hydrogen, toward \wkk\  of $4.34\times10^{21}\pscm$ \citep{Willingale} and appropriate abundances \citep{Wilms} are adopted in all spectral fits. All final models will require addition neutral absorption on the level of $\sim10^{21}\pscm$ that is local to \wkk.

Fitting a power law  to the spectra between $4-5$, $7.5-10$, and $40-50\keV$, and extrapolated over the entire band reveals an excess of emission in the \feka\ band ($6-7\keV$) and between $10-40\keV$ (Fig.~\ref{fig:spec}). However, unlike most Seyfert 1 galaxies, there is no soft-excess evident below $\sim2\keV$, but rather a dearth.  This is likely from additional absorption local to \wkk.

A Gaussian profile with free energy and width ($\sigma$) is substantially better than a narrow ($\sigma=1\eV$) profile with the energy fixed at $E=6.4\keV$ ($\delC=-27$ for 2 additional parameters).  In this case, the centroid energy is $6.20^{+0.20}_{-0.36}\keV$ and the width is $\sigma=0.67^{+0.41}_{-0.26}\keV$.  Assuming that the width of the line is attributed to velocity broadening from material in a Keplerian orbit, the emission would originate at a distance of $66^{+39}_{-51}\rg$ from the black hole. 

The absorption features reported by \cite{JiangWKK} and attributed to an ultrafast outflow with argon and iron over-abundances are less conspicuous in the 2018 data (Fig.~\ref{fig:spec}).  Applying the same absorption grid used in \cite{JiangWKK} with all wind parameters allowed to vary does not improve the fit significantly.  The feature previously attributed to iron is negligible in the new data and now results in sub-solar iron abundances.  Alternatively, fixing all wind parameters to the values found in \cite{JiangWKK} except for the column density finds a value of $\ls7\times10^{22}\pscm$.  This is comparable to $\sim12\times10^{22}\pscm$ found by \cite{JiangWKK}.  Variability of the ultrafast outflow is possible and could explain the differences in appearance with the earlier \suzaku\ and \nustar\ observations \citep{JiangWKK}.  However, the features are not considered further here and it is noted that the continuum measurements found here are comparable to the work of \cite{JiangWKK}, which included a wind component in their model. Therefore, accounting for those features does not significantly impact the continuum results.

Introducing a {\sc xillverCp}  \citep{xillver} component to model distant, neutral reflection along with additional host-galaxy absorption provided a reasonable fit ($\Cdof=1012.5/898$, where dof is the degrees of freedom, and $\rm DIC=1022.6$).  This will serve as the baseline model to compare subsequent fits.  The apparent absence of a soft-excess, which could be from an excess of absorption and/or from a weak soft component, is unusual for a Seyfert~1.  To examine for a soft-excess,  multi-component continuum models are attempted. 
Following \cite{Petrucci18}, the primary continuum is modelled with two {\sc nthComp} \citep{nthComp1, nthComp2} components to replicate two coronae (e.g. \citealt{wc1,wc2,Ballantyne20}).  The warm and hot coronae are each allowed to have their own electron temperatures ($kT_e^{wc}$ and $kT_e^{hc}$, respectively), but share a common black body disc temperature ($kT_{bb}=0.05\keV$).  The additional host-galaxy column density and a distant reflector remain in the model.  During initial fits, the temperature of the warm corona that accounts for the putative soft-excess was not constrained.  Consequently, this parameter was fixed to a typical value of $kT_e^{wc}=1\keV$ \citep{Petrucci18}.
The double Comptonisation model attaches a soft-excess component in the intrinsic X-ray spectrum of \wkk, which may be more plausible, but the fit is statistically less likely than the baseline single power law and distant, neutral reflector ($\delDIC=+5.7$). 

Alternatively, a natural explanation for the excess emission at  $\sim20\keV$ and in the \feka\ band, which is broad and redward of the neutral line, is relativistically blurred reflection from the inner disc (e.g. \citealt{RossFabian05}).  This model also accounts for fluorescent emission from lower atomic number elements that contribute to forming a so-called soft excess.  The distant reflector is replaced with blurred reflection using the {\sc relxill} suite of models \citep{relline, xillver}.  Initially, several flavours of {\sc relxill} were tested with various combinations of free parameters.  Allowing for variable densities or coronal electron temperatures never improved the fits so these parameters are fixed to their canonical values of $N=10^{15}\pccm$ and $kT_e =100\keV$, respectively.  In addition, the black hole spin, disc inclination ($i$), and iron abundance ($\rm A_{Fe}$) are linked between epochs.

Two reasonable blurred reflection fits were achieved using {\sc relxillD} ($\Cdof=993.1/890$ and $\rm DIC=1011.1$) and {\sc relxilllp} ($\Cdof=1004.1/891$ and $\rm DIC=1025.8$), separately.  With the former, the emissivity profile for the accretion disc is described by a broken power law, where the profile goes as $r^{-q_{in}}$ from the inner disc edge ($R_{in}$) to the break radius $R_b$, where it then changes to $r^{-q_{out}}$.  Beyond the break radius of $R_b$, the outer emissivity index is set to the classical value of $q_{out} = 3$.  Such a model depicting differing illumination patterns could be indicative of different coronal geometries (e.g. \citealt{WF12}).

With {\sc relxilllp}, the corona is assumed to be a point source that is located on the spin axis at some height above the disc (i.e. a lamp-post). Here, the height of the corona point source is the model parameter and the illumination pattern on the disc (i.e. the emissivity profile) is determined from that.  There have been significant advances in spectral modelling (e.g. \citealt{Dauser13, WG15, Gonzalez17, Jiang22}) and reverberation mapping (e.g. \citealt{Caballero18, Caballero20, Alston20}) that have enabled good measurements of the coronal height in other AGN.  Both spectral and timing methods provide comparable estimates when applied to the same source.

In both reflection models presented, the fundamental assumption is that the inner disc edge ($R_{in}$) extends down to the innermost stable circular orbit (ISCO), which is driven by the angular momentum of the black hole defined by the spin parameter ($a_*$).

\begin{figure}
   \centering
   {\scalebox{0.32}{\includegraphics[trim= 2.5cm 1cm 1cm 0cm, angle=-90,clip=true]{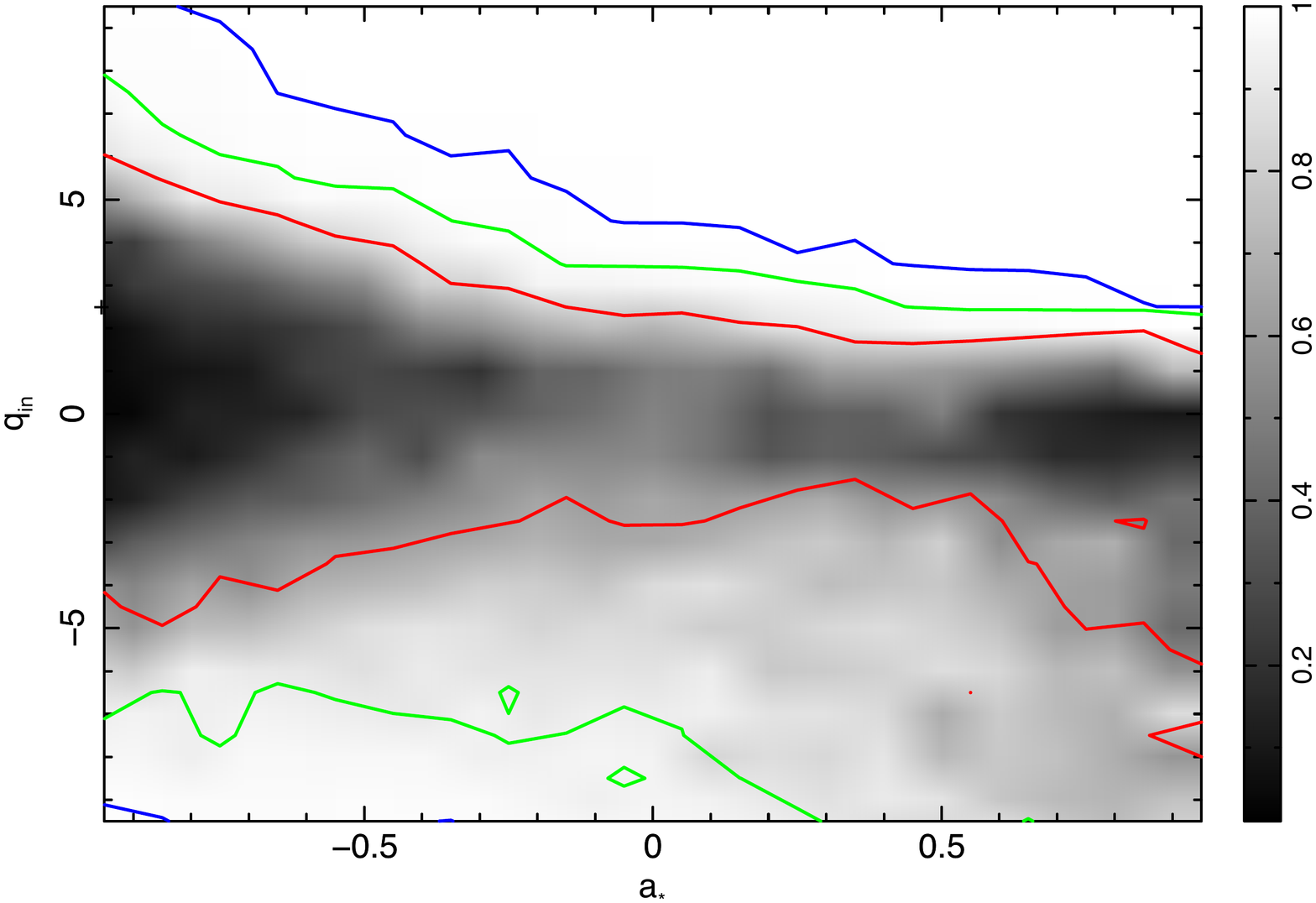}}}
   {\scalebox{0.32}{\includegraphics[trim= 2.5cm 1cm 1cm 0cm, angle=-90, clip=true]{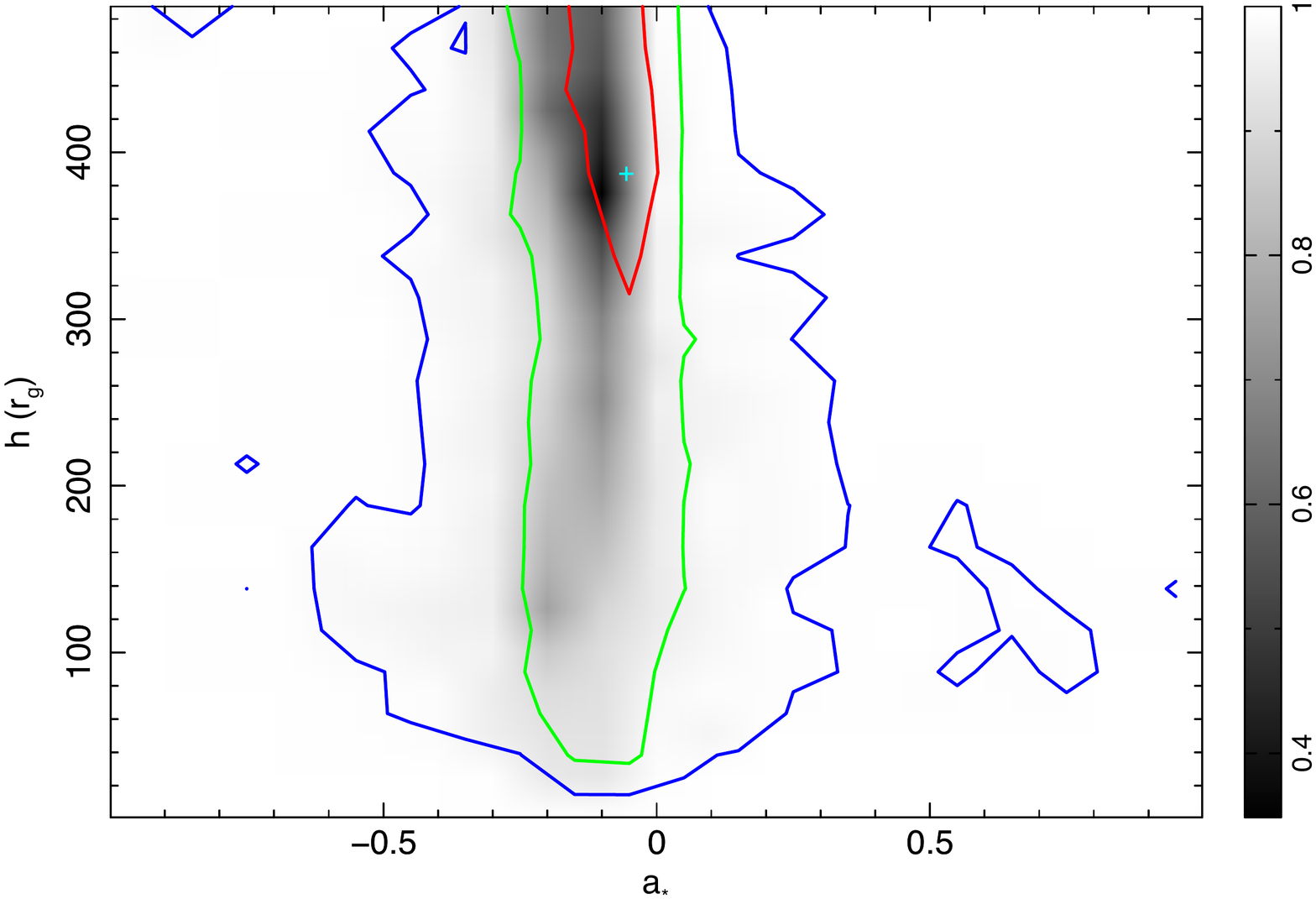}}}
   \caption{The integrated probability density contours assuming the inner disc radius extends down to the innermost stable circular orbit renders the following behaviour between some parameters. {\bf Upper panel:} The inner emissivity index and black hole spin are both consistent with low values though neither is well-constrained.    {\bf Lower panel:} In the lamp-post scenario, the spin is tightly constrained  near zero, and large values for the height of the point source corona are preferred. The contour levels depicted are 0.68, 0.95, and 0.997.  }
   \label{fig:odd}
\end{figure}

While both models provide reasonable fits to the spectra, the fits do have some unusual characteristics.  For the {\sc relxillD} interpretation, though both the inner-disc emissivity profile and black hole spin are poorly constrained, the best-fit values are both small or negative (Fig.~\ref{fig:odd}).  This would indicate that the disc brightens with increasing distance and the black hole spin is retrograde.  For the lamp-post model, the best-fit height parameter suggests that the point source corona is at $h>128\rg$, corresponding to a light-travel time of over 30-minutes for a $2\times10^6\Msun$ black hole \citep{Malizia08}.  The black hole spin is well constrained in this model to be retrograde and near zero ($a_*=-0.09^{+0.03}_{-0.17}$).

The combination of parameters measured with these models are atypical.  Notably, is the low spin parameter.  Nearly all of the three-dozen black hole spin parameters measured in active galaxies are consistent with high spins ($a_*>0.9$; e.g. \citealt{Reynolds21}).   There are several reasons for this including cosmological (e.g. \citealt{Berti08}), selection effects (e.g. \citealt{Reynolds21}) and the general difficulty with measuring low-spins robustly from limited bandpass data (e.g. \citealt{BG16}).   For those reasons, the same two blurred reflection models were applied to \wkk, but without the fundamental assumption that the inner disc edge reaches the ISCO.  Rather, the black hole spin parameter is fixed to the maximum value ($a_*=0.998$, \citealt{Thorne74}) and the inner accretion disc radius is permitted to vary freely.   Compared to the models with variable spin, the model with fixed maximum spin and free inner disc radius was comparably good  with {\sc relxillD}  ($\delDIC=+2.8$) and significantly better in the {\sc relxilllp} case ($\delDIC=-10.0$). 

\begin{figure}
\begin{center}
\scalebox{0.33}{\includegraphics[trim= 1cm 1.5cm 2cm 3cm, angle=0,clip=true]{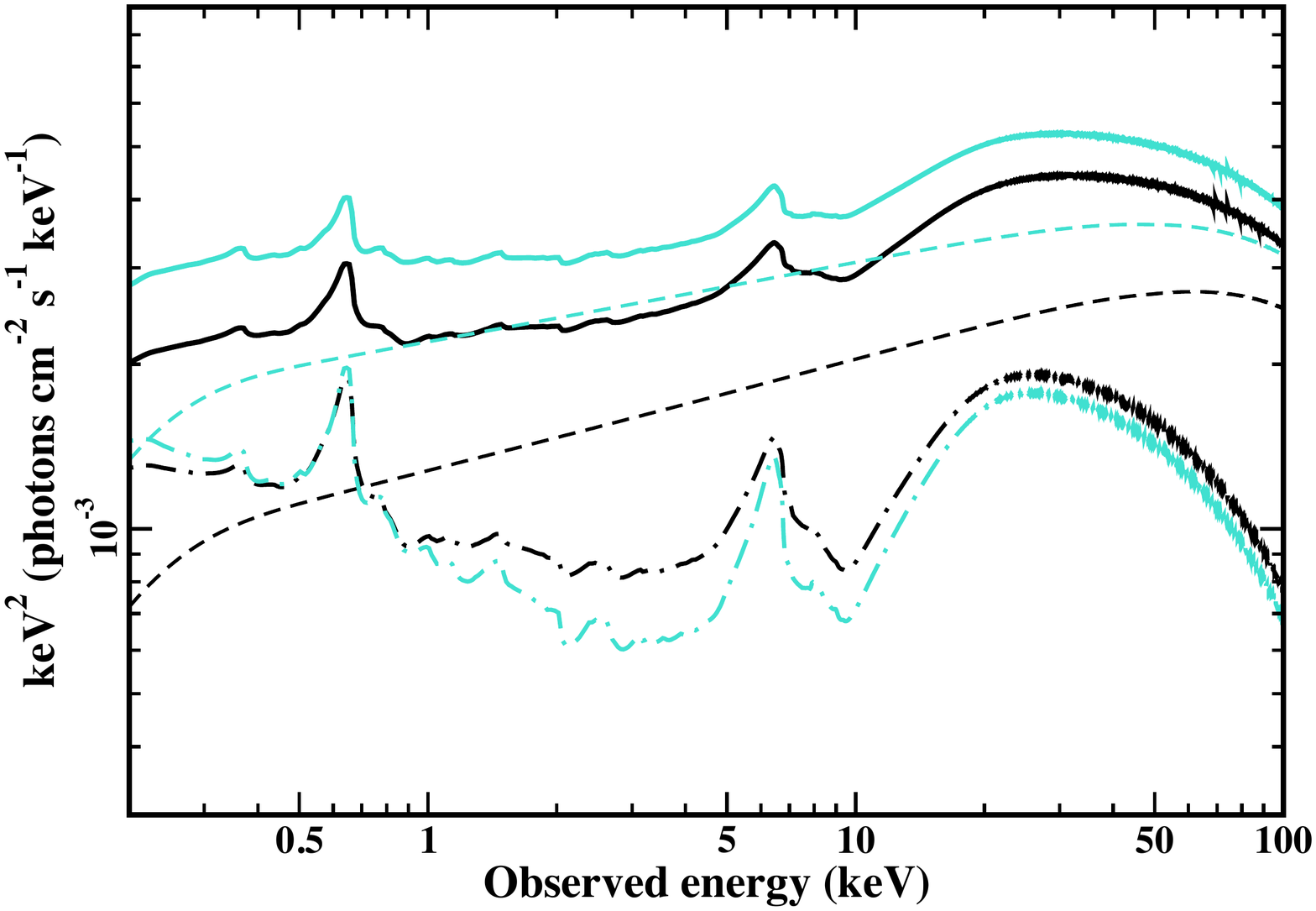}}
\scalebox{0.33}{\includegraphics[trim= 1cm 1.5cm 2cm 5cm, clip=true,angle=0]{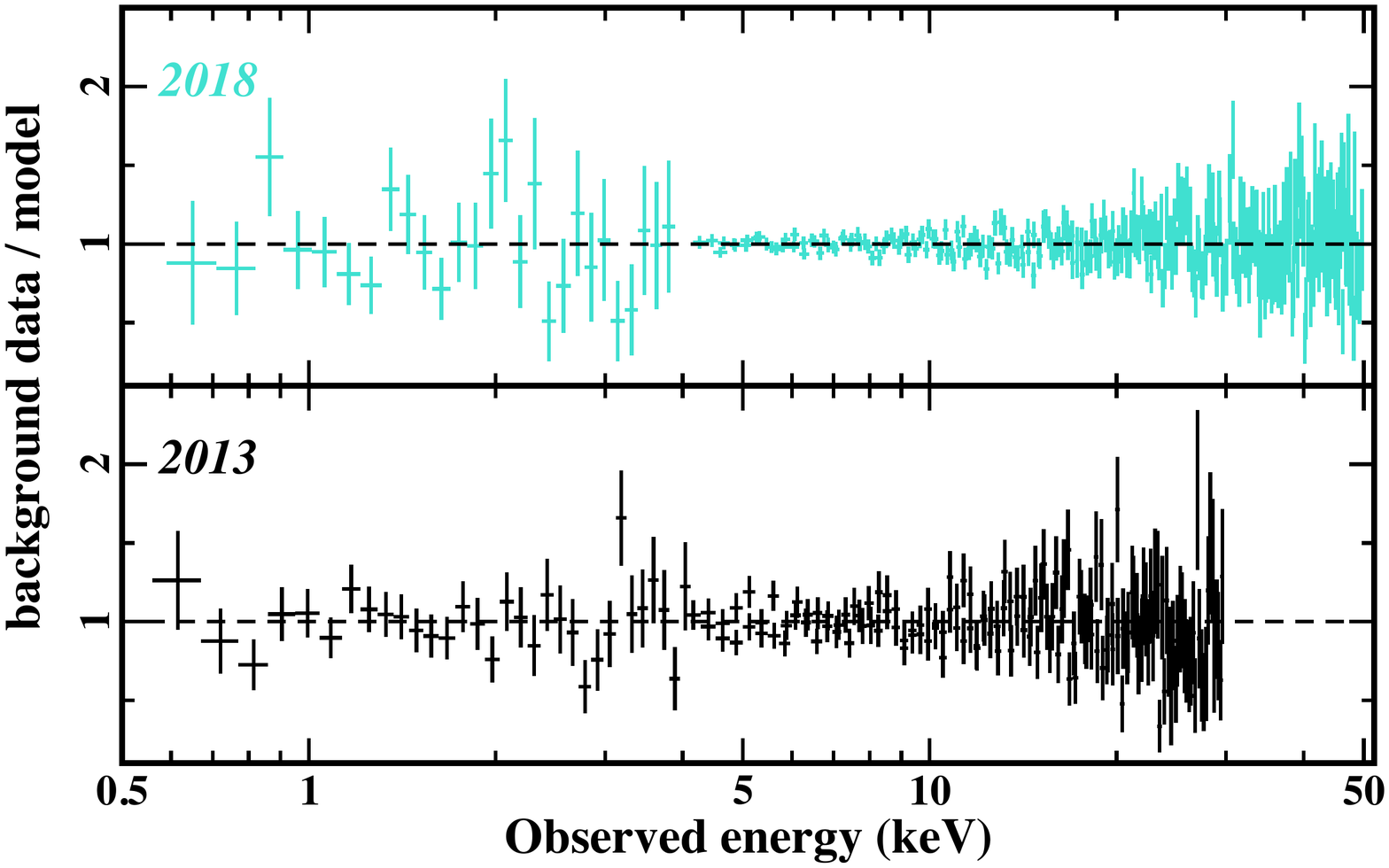}}
\end{center}
\caption{{\bf Upper panel:} The components of the best fitting blurred reflection model (Table~\ref{tab:fit}) with the Galactic and host column density removed.  The solid curves are the total models (black for 2013 and turquoise for 2018).  The power law and blurred reflector are shown as dashed curves and dot-dash curves, respectively.    {\bf Lower panels:} The data-to-model residuals remaining in the spectrum of each epoch. 
}
\label{fig:fit}
\end{figure}

\begin{figure}
   \centering
   {\scalebox{0.33}{\includegraphics[trim= 1.2cm 3cm 3cm 1cm, angle=90,clip=true]{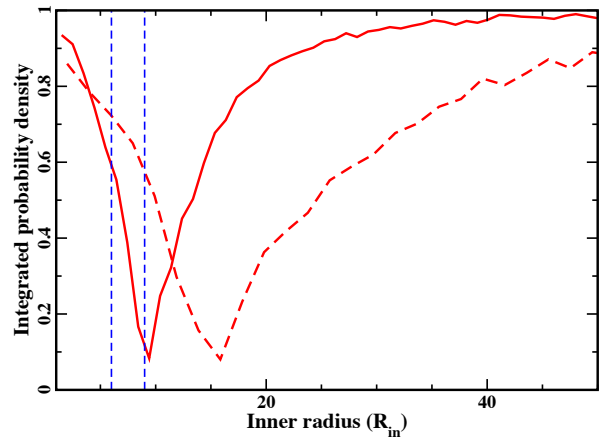}}}
   \caption{Assuming a Kerr black hole with  the inner disc radius truncated beyond $\risco$ provides a better statistical fit with more plausible parameters.   The constraints on the inner disc radius for the {\sc relxillD} (solid red curve)  and {\sc relxilllp} (dashed red curve) are shown.  The parameter is better constrained with {\sc relxillD} and both models favour an inner disc that does not extend to the ISCO.  The blue vertical lines mark the radii expected in a Schwarzschild black hole ($6\rg$) and one with maximum retrograde spin ($9\rg$).
   }
   \label{fig:delC}
\end{figure}

For the {\sc relxillD} model, in addition to $R_{in}$, the inner emissivity profile $q_{in}$ was also allowed to vary.  This did not enhance the fit or constrain the parameter, basically mimicking the behaviour in Fig.~\ref{fig:odd}.  For simplicity, in the final fit we assumed $q=3$ across the entire disc commensurate with a Newtonian system.  This generated a good fit ($\Cdof=995.5/892$ and $\rm DIC=1013.9$) and while statically it is slightly less preferred than the scenario with $R_{in}=\risco$ ($\delDIC=+2.8$), the model parameters seem more plausible (Fig.~\ref{fig:fit} and Table~\ref{tab:fit}) in our view since low/retrograde spins and inverted emissivity profiles are not required.   In this scenario, the inner disc radius is relatively well constrained to $9.9^{+20.1}_{-5.4}\rg$ thus having the disc extending down to the ISCO of a Kerr black hole ($\risco=1.25\rg$) is unlikely (Fig.~\ref{fig:delC}).

The {\sc relxilllp} model with a Kerr black hole and free $R_{in}$ also produces a good fit ($\Cdof=997.8/891$ and $\rm DIC=1015.8$) that is comparable to the  {\sc relxillD} model with free $R_{in}$ ($\delDIC=+1.9$). Here the inner radius is poorly constrained, but the best-fit is with $R_{in}\sim15\rg$, similar to the {\sc relxillD} case.  The height of the corona is well constrained to be $7.5^{+5.1}_{-2.5}\rg$ unlike how poorly constrained the height was in the {\sc relxilllp} model with variable spin (Fig.~\ref{fig:odd}).  It is notable that leaving $R_{in}$ free to vary or tying it to the ISCO and changing $a_*$ actually results in comparable values for the location of the inner edge.  
\begin{table}
\caption{The blurred reflection model fitted to the 2018 and 2013 \wkk\ spectra.  
The model components and model parameters are listed in Columns 1 and 2, respectively. 
Columns 3 and 4 list the parameter values during the 2018 and 2013 observations, respectively.
The photon index of the reflection components is linked to the photon index measured by {\sc nthComp}.
The reflection fraction ($\mathcal{R}$) is calculated as the ratio of the reflected flux over the power law flux in the $0.1-100\keV$ band and is based on the best-fit parameters.  Setting $\mathcal{R}$ as a fit parameter in the {\sc relxillD} model generates similar values.  The cutoff temperature in {\sc relxillD} is fixed at $E_{c}=300\keV$.
Values that are linked between epochs appear only in column 3.  
}
\centering
\scalebox{0.92}{
\begin{tabular}{ccccc}                
\hline
(1) & (2) & (3) & (4)   \\
 Model  &  Model   &  2018 & 2013 \\
  Component &   Parameter  &   &  \\
\hline
 Neutral  		  & $\nh$ 					& $1.78^{+0.86}_{-0.75}$ 		&   \\
 absorption  & ($\times10^{21}\pscm$) & \\
  ({\sc ztbabs}) & & \\
\hline
 Power law 		  & $\Gamma$ 					& $1.87^{+0.13}_{-0.08}$ 		& $1.80^{+0.10}_{-0.12}$   \\
 ({\sc nthComp})       & $kT_e /\keV$ 			        & $100^{f}$ 			&  \\
  			          & $kT_{bb} /\keV$ 			& $0.05^{f}$ 		&  \\
        			      &  norm ($\times10^{-3}$)		        & $2.20^{+0.98}_{-0.59}$  & $1.29^{+0.73}_{-0.92}$\\
\hline
 Blurred      & $q$                                       & $3^f$         			 &   \\
reflection          & $R_{in}/\rg$   			       & $9.9^{+20.1}_{-5.4}$     		         &                                \\
      ({\sc relxillD})                            & $R_{out} /\rg$	               & $400^{f}$    			 &                                \\
       				 & $a_*$  			                        & $0.998^{f}$	 &   			\\
       				 & $i /\deg$		                         & $24^{+14}_{-15}$		         &                                      \\
       				 & $\Gamma$ 					& $1.87$ 		                  & $1.80$   \\
      				 & log($\xi/\erg\cmps$)                      & $2.98^{+0.32}_{-0.95}$  	& $3.11^{+0.48}_{-0.47}$       \\
       				 & $\rm A_{Fe} / A_{\odot}$                & $1.00^{+1.94}_{-0.22}$         \\
       				 & log($N/\pccm$)                             & $15^{f}$  &      \\
        				 &  norm ($\times10^{-5}$)		        & $1.09^{+0.56}_{-0.26}$  & $1.07^{+0.53}_{-0.40}$   \\
  			        & $\mathcal{R}$                                 & $0.46^{+0.32}_{-0.17}$                              & $0.75^{+0.57}_{-0.60}$  \\
\hline
 Calibration   & \nustar\ 		         & $0.99\pm0.02$ &  		$0.94\pm0.04$	          \\
 constant      & \swift\ 	& $0.92\pm0.12$   & $0.97\pm0.10$\\
\hline
          Fit Quality 	& $\Cdof$						 & $995.5/892$              &    \\
           			& DIC    						 & $1013.9$ 		       &      \\
\hline
\label{tab:fit}
\end{tabular}
}
\end{table}

\section{Discussion \& Conclusion} 
\label{sect:dis}

\wkk\ exhibits rapid flux variability on short time scales, but spectral variations appear minimal. Two \nustar\ observations from 2013 and 2018 are examined here, and both are relatively consistent save for a modest change in brightness.  The photon indices measured here are also comparable to the earlier 2006 and 2007 \swift\ XRT data \citep{Malizia08}.

The $0.5-50\keV$ spectra at both epochs can be described well with a blurred reflection scenario where some of the primary emission from the corona illuminates the inner accretion disc and is backscattered in the direction of the observer (e.g. \citealt{RossFabian05}).  Applying the reflection models with the standard assumption that the inner edge of the accretion disc extends down to the ISCO results in an unexpected combination of parameter values and requires an unconventional interpretation.  

For example, when adopting a {\sc relxill} scenario, the black hole spin parameter and inner emissivity index are poorly constrained, but positive spin values ($a_*>0$) in combination with steep emissivity profiles ($q_{in}>5$), as might be expected, are significantly ruled out.  Instead, high spin values are better paired with inverted emissivity profiles, and conversely, low (retrograde) spins with higher $q_{in}$ (Fig.~\ref{fig:odd}).   Both of these conditions imply that the inner disc emission close to the black hole is weak -- either the inner disc is located at large distances because the black hole spins slowly or the disc is brighter at larger distances.   Similarly, the lamp-post scenarios ({\sc relxilllp}) find a large height, that is poorly constrained, for the point source corona (Fig.~\ref{fig:odd}), but tightly constrain the spin parameter to $a_*\approx0$.

However, models assuming that \wkk\ possesses a maximum spinning black hole with its inner disc truncated at larger distances, generate similar (or better) fits and more sensible parameters.  In these cases, the inner disc does not extend down to the ISCO, but is truncated at $5-20\rg$.  The flatter emissivity profiles ($q=3$) and low reflections fractions ($\mathcal{R}<1$) that are found in these models follow from the interpretation that the accretion disc inner edge might not extend into the immediate vicinity of the black hole where the  general relativistic effects that produce large $\mathcal{R}$ and $q$ are at work.  With these maximum-spin models, the measured corona height in \wkk\ is also better constrained and comparable to the heights measured in other systems (e.g. \citealt{Caballero18, Caballero20, Alston20, Gonzalez20, G15}).

One could also have a situation where the lamp-post is at some horizontal distance and orbits the rotation axis.  This might enhance the disc brightness at larger distances, but the inner edge should still be detectable (e.g. \citealt{WF12, Gonzalez17}).  This scenario would not change the measured inner radius seen here, which appears independent of the black hole spin. 

During the fainter stages of the hard state of black hole binaries, truncated discs are common (e.g. \citealt{Esin97}).  For supermassive black holes, such discs are typically associated with LLAGN where the standard disc \citep{Shakura73} is truncated at $>100\rg$ and the inner flow is replaced with a hot, radiative inefficient flow (e.g. \citealt{ADAF, Lasota}) that is extremely sub-Eddington ($L/L_{Edd}<10^{-3}$).  

\wkk\ does not exhibit the characteristics of a LLAGN.   Following \cite{Brightman13}, based on the measured photon indices in 2013 and 2018, $L/L_{Edd}=0.03-0.06$.  Even though the luminosity ratio is not particularly high, the value is consistent with a standard accretion disc.   Its UV bump is probably present given the estimated UV-to-X-ray spectral slope of $\alpha_{ox}\sim-1.40$ \citep{Panessa11} and \wkk\ does not appear to be radio-loud or exhibit synchrotron emission related to jets.   It is not detected in the TIFR GMRT Sky Survey (TGSS, \citealt{TGSS})  and has a $3\sigma$ upper-limit of $<10$~mJy at 150~MHz.  Nor is it detected at $72-231$~MHz in the GLEAM survey \citep{GLEAM}.   \wkk\ is also highly variable in the X-rays, contrary to most LLAGN  (e.g.  \citealt{Ptak98, Younes19}) and it possesses a broad \feka\ line and Compton hump, which are not normally seen in LLAGN (e.g. \citealt{Dewangan04, Ptak04, Reynolds09, Lobban10, Younes19}). 

Noteworthy, \wkk\ is categorized as a narrow-line Seyfert 1 galaxy (NLS1) by \cite{Masetti06}.  Only an upper-limit could be measured for the \feii\ emission, but in conjunction with the other optical properties it does meet the criteria.  This is of interest because the class includes some extreme objects that are well known for emitting X-rays from a compact region close to a Kerr black hole (e.g. \citealt{Fabian09, Jiang18, Wilkins22}; see \citealt{G18} for review).  This appears contradictory to the behaviour of \wkk, whose inner disc is at approximately $10\rg$.  It is worth noting that \wkk\ was anomalous compared to other NLS1s observed with \suzaku\ \citep{Waddell20, Waddell22}.  Compared to other NLS1s in the \suzaku\ sample, \wkk\ possesses relatively weak reflection features and soft excess, and a low Eddington luminosity ratio, which are more consistent with typical Seyfert 1 galaxies.  The contradiction may speak to the need for a better definition of the NLS1 phenomenon. 

If the corona is covering a large fraction of the inner disc, the reflection spectrum could be highly Comptonised (e.g. \citealt{WG15Comp, Petrucci01}) making the inner disc appear truncated.  This does not seem applicable in \wkk\ as the continuum spectrum is rather typical in shape.  A cursory application of such a model ({\sc comptonise}; \citealt{WG15Comp}) to the data shows that the covering fraction is rather low ($\sim5$ per cent) and the reflection parameters are not dissimilar to the previous fits in this work. 

\wkk\ is relatively under-examined.  The \nustar\ data presented here are the highest quality to date, but do not extend below $\sim4\keV$.  The host galaxy exists in an optically crowded field (see figure 2 in the Online Material of \citealt{Masetti06})  and has not been investigated significantly.  Perhaps a recent merger could have disrupted the disc and it is now refilling.  The line-of-sight is also moderately absorbed by the Galaxy and the host ($\sim10^{21}\pscm$) making inspection of the soft band more challenging.  Understanding if and when the disc does extend to the ISCO is important since it is the  fundamental assumption behind measuring the black hole spin (e.g. \citealt{Shafee06, Brenneman06}; see \citealt{Reynolds21} for a recent review).  It is necessary to determine how often this assumption may be invalid.  

Our analysis suggests that \wkk\ might have a low or retrograde spinning black hole, but the more likely possibility is that the disc is truncated and the black hole spins rapidly.  Future observations with \xmm\ and \nustar\ will constrain the soft X-ray emission and search for frequency-dependent lags. Studies of the host galaxy will confirm the behaviour and origin of the truncated disc in \wkk.


\section*{Acknowledgments}
The authors thank the referee for comments that improved the manuscript.
This research has made use of data obtained with NuSTAR, a project led by Caltech, funded by NASA and managed by NASA/JPL.  
LCG acknowledge financial support from the Natural Sciences and Engineering Research Council of Canada (NSERC) and from the Canadian Space Agency (CSA).  JJ acknowledges support from the Leverhulme Trust, the Isaac Newton Trust and St Edmund's College, University of Cambridge.

\section*{Data Availability}
The data used in this study are available in the \nustar\ and \swift\ public archives.


\bibliographystyle{mnras}
\bibliography{bibfile} 

\begin{thebibliography}{}
\makeatletter
\relax
\def\mn@urlcharsother{\let\do\@makeother \do\$\do\&\do\#\do\^\do\_\do\%\do\~}
\def\mn@doi{\begingroup\mn@urlcharsother \@ifnextchar [ {\mn@doi@}
  {\mn@doi@[]}}
\def\mn@doi@[#1]#2{\def\@tempa{#1}\ifx\@tempa\@empty \href
  {http://dx.doi.org/#2} {doi:#2}\else \href {http://dx.doi.org/#2} {#1}\fi
  \endgroup}
\def\mn@eprint#1#2{\mn@eprint@#1:#2::\@nil}
\def\mn@eprint@arXiv#1{\href {http://arxiv.org/abs/#1} {{\tt arXiv:#1}}}
\def\mn@eprint@dblp#1{\href {http://dblp.uni-trier.de/rec/bibtex/#1.xml}
  {dblp:#1}}
\def\mn@eprint@#1:#2:#3:#4\@nil{\def\@tempa {#1}\def\@tempb {#2}\def\@tempc
  {#3}\ifx \@tempc \@empty \let \@tempc \@tempb \let \@tempb \@tempa \fi \ifx
  \@tempb \@empty \def\@tempb {arXiv}\fi \@ifundefined
  {mn@eprint@\@tempb}{\@tempb:\@tempc}{\expandafter \expandafter \csname
  mn@eprint@\@tempb\endcsname \expandafter{\@tempc}}}

\bibitem[\protect\citeauthoryear{{Alston} et~al.,}{{Alston}
  et~al.}{2020}]{Alston20}
{Alston} W.~N.,  et~al., 2020, \mn@doi [Nature Astronomy]
  {10.1038/s41550-019-1002-x}, \href
  {https://ui.adsabs.harvard.edu/abs/2020NatAs...4..597A} {4, 597}

\bibitem[\protect\citeauthoryear{{Ballantyne}}{{Ballantyne}}{2020}]{Ballantyne20}
{Ballantyne} D.~R.,  2020, \mn@doi [\mnras] {10.1093/mnras/stz3294}, \href
  {https://ui.adsabs.harvard.edu/abs/2020MNRAS.491.3553B} {491, 3553}

\bibitem[\protect\citeauthoryear{{Berti} \& {Volonteri}}{{Berti} \&
  {Volonteri}}{2008}]{Berti08}
{Berti} E.,  {Volonteri} M.,  2008, \mn@doi [\apj] {10.1086/590379}, \href
  {https://ui.adsabs.harvard.edu/abs/2008ApJ...684..822B} {684, 822}

\bibitem[\protect\citeauthoryear{{Bonson} \& {Gallo}}{{Bonson} \&
  {Gallo}}{2016}]{BG16}
{Bonson} K.,  {Gallo} L.~C.,  2016, \mn@doi [\mnras] {10.1093/mnras/stw466},
  \href {https://ui.adsabs.harvard.edu/abs/2016MNRAS.458.1927B} {458, 1927}

\bibitem[\protect\citeauthoryear{{Brenneman} \& {Reynolds}}{{Brenneman} \&
  {Reynolds}}{2006}]{Brenneman06}
{Brenneman} L.~W.,  {Reynolds} C.~S.,  2006, \mn@doi [\apj] {10.1086/508146},
  \href {https://ui.adsabs.harvard.edu/abs/2006ApJ...652.1028B} {652, 1028}

\bibitem[\protect\citeauthoryear{{Brightman} et~al.,}{{Brightman}
  et~al.}{2013}]{Brightman13}
{Brightman} M.,  et~al., 2013, \mn@doi [\mnras] {10.1093/mnras/stt920}, \href
  {https://ui.adsabs.harvard.edu/abs/2013MNRAS.433.2485B} {433, 2485}

\bibitem[\protect\citeauthoryear{{Burrows} et~al.,}{{Burrows}
  et~al.}{2005}]{xrt}
{Burrows} D.~N.,  et~al., 2005, \mn@doi [\ssr] {10.1007/s11214-005-5097-2},
  \href {https://ui.adsabs.harvard.edu/abs/2005SSRv..120..165B} {120, 165}

\bibitem[\protect\citeauthoryear{{Caballero-Garc{\'\i}a}, {Papadakis},
  {Dov{\v{c}}iak}, {Bursa}, {Epitropakis}, {Karas}  \&
  {Svoboda}}{{Caballero-Garc{\'\i}a} et~al.}{2018}]{Caballero18}
{Caballero-Garc{\'\i}a} M.~D.,  {Papadakis} I.~E.,  {Dov{\v{c}}iak} M.,
  {Bursa} M.,  {Epitropakis} A.,  {Karas} V.,   {Svoboda} J.,  2018, \mn@doi
  [\mnras] {10.1093/mnras/sty1990}, \href
  {https://ui.adsabs.harvard.edu/abs/2018MNRAS.480.2650C} {480, 2650}

\bibitem[\protect\citeauthoryear{{Caballero-Garc{\'\i}a}, {Papadakis},
  {Dov{\v{c}}iak}, {Bursa}, {Svoboda}  \& {Karas}}{{Caballero-Garc{\'\i}a}
  et~al.}{2020}]{Caballero20}
{Caballero-Garc{\'\i}a} M.~D.,  {Papadakis} I.~E.,  {Dov{\v{c}}iak} M.,
  {Bursa} M.,  {Svoboda} J.,   {Karas} V.,  2020, \mn@doi [\mnras]
  {10.1093/mnras/staa2554}, \href
  {https://ui.adsabs.harvard.edu/abs/2020MNRAS.498.3184C} {498, 3184}

\bibitem[\protect\citeauthoryear{{Cash}}{{Cash}}{1979}]{Cash}
{Cash} W.,  1979, \mn@doi [\apj] {10.1086/156922}, \href
  {https://ui.adsabs.harvard.edu/abs/1979ApJ...228..939C} {228, 939}

\bibitem[\protect\citeauthoryear{{Czerny}, {Niko{\l}ajuk},
  {R{\'o}{\.z}a{\'n}ska}, {Dumont}, {Loska}  \& {Zycki}}{{Czerny}
  et~al.}{2003}]{wc2}
{Czerny} B.,  {Niko{\l}ajuk} M.,  {R{\'o}{\.z}a{\'n}ska} A.,  {Dumont} A.~M.,
  {Loska} Z.,   {Zycki} P.~T.,  2003, \mn@doi [\aap]
  {10.1051/0004-6361:20031441}, \href
  {https://ui.adsabs.harvard.edu/abs/2003A&A...412..317C} {412, 317}

\bibitem[\protect\citeauthoryear{{Dauser}, {Wilms}, {Reynolds}  \&
  {Brenneman}}{{Dauser} et~al.}{2010}]{relline}
{Dauser} T.,  {Wilms} J.,  {Reynolds} C.~S.,   {Brenneman} L.~W.,  2010,
  \mn@doi [\mnras] {10.1111/j.1365-2966.2010.17393.x}, \href
  {https://ui.adsabs.harvard.edu/abs/2010MNRAS.409.1534D} {409, 1534}

\bibitem[\protect\citeauthoryear{{Dauser}, {Garcia}, {Wilms}, {B{\"o}ck},
  {Brenneman}, {Falanga}, {Fukumura}  \& {Reynolds}}{{Dauser}
  et~al.}{2013}]{Dauser13}
{Dauser} T.,  {Garcia} J.,  {Wilms} J.,  {B{\"o}ck} M.,  {Brenneman} L.~W.,
  {Falanga} M.,  {Fukumura} K.,   {Reynolds} C.~S.,  2013, \mn@doi [\mnras]
  {10.1093/mnras/sts710}, \href
  {https://ui.adsabs.harvard.edu/abs/2013MNRAS.430.1694D} {430, 1694}

\bibitem[\protect\citeauthoryear{{Dewangan}, {Griffiths}, {Di Matteo}  \&
  {Schurch}}{{Dewangan} et~al.}{2004}]{Dewangan04}
{Dewangan} G.~C.,  {Griffiths} R.~E.,  {Di Matteo} T.,   {Schurch} N.~J.,
  2004, \mn@doi [\apj] {10.1086/383607}, \href
  {https://ui.adsabs.harvard.edu/abs/2004ApJ...607..788D} {607, 788}

\bibitem[\protect\citeauthoryear{{Esin}, {McClintock}  \& {Narayan}}{{Esin}
  et~al.}{1997}]{Esin97}
{Esin} A.~A.,  {McClintock} J.~E.,   {Narayan} R.,  1997, \mn@doi [\apj]
  {10.1086/304829}, \href
  {https://ui.adsabs.harvard.edu/abs/1997ApJ...489..865E} {489, 865}

\bibitem[\protect\citeauthoryear{Evans et~al.,}{Evans et~al.}{2009}]{Evans09}
Evans P.~A.,  et~al., 2009, \mn@doi [Monthly Notices of the Royal Astronomical
  Society] {10.1111/j.1365-2966.2009.14913.x}, 397, 1177

\bibitem[\protect\citeauthoryear{{Fabian} et~al.,}{{Fabian}
  et~al.}{2009}]{Fabian09}
{Fabian} A.~C.,  et~al., 2009, \mn@doi [\nat] {10.1038/nature08007}, \href
  {https://ui.adsabs.harvard.edu/abs/2009Natur.459..540F} {459, 540}

\bibitem[\protect\citeauthoryear{{Gallo}}{{Gallo}}{2018}]{G18}
{Gallo} L.,  2018, in Revisiting Narrow-Line Seyfert 1 Galaxies and their Place
  in the Universe. p.~34 (\mn@eprint {arXiv} {1807.09838})

\bibitem[\protect\citeauthoryear{{Gallo} et~al.,}{{Gallo} et~al.}{2015}]{G15}
{Gallo} L.~C.,  et~al., 2015, \mn@doi [\mnras] {10.1093/mnras/stu2108}, \href
  {https://ui.adsabs.harvard.edu/abs/2015MNRAS.446..633G} {446, 633}

\bibitem[\protect\citeauthoryear{{Gammie}, {Narayan}  \& {Blandford}}{{Gammie}
  et~al.}{1999}]{Gammie99}
{Gammie} C.~F.,  {Narayan} R.,   {Blandford} R.,  1999, \mn@doi [\apj]
  {10.1086/307089}, \href
  {https://ui.adsabs.harvard.edu/abs/1999ApJ...516..177G} {516, 177}

\bibitem[\protect\citeauthoryear{{Garc{\'\i}a} \& {Kallman}}{{Garc{\'\i}a} \&
  {Kallman}}{2010}]{xillver}
{Garc{\'\i}a} J.,  {Kallman} T.~R.,  2010, \mn@doi [\apj]
  {10.1088/0004-637X/718/2/695}, \href
  {https://ui.adsabs.harvard.edu/abs/2010ApJ...718..695G} {718, 695}

\bibitem[\protect\citeauthoryear{{Gehrels} et~al.,}{{Gehrels}
  et~al.}{2004}]{swift}
{Gehrels} N.,  et~al., 2004, \mn@doi [\apj] {10.1086/422091}, \href
  {https://ui.adsabs.harvard.edu/abs/2004ApJ...611.1005G} {611, 1005}

\bibitem[\protect\citeauthoryear{Gelman, Carlin, Stern  \& Rubin}{Gelman
  et~al.}{2003}]{dic2}
Gelman A.,  Carlin J.,  Stern H.,   Rubin D.,  2003, Bayesian Data Analysis,
  Second Edition.
Chapman \& Hall/CRC Texts in Statistical Science, Taylor \& Francis, \url
  {https://books.google.ca/books?id=TNYhnkXQSjAC}

\bibitem[\protect\citeauthoryear{{Gonzalez}, {Wilkins}  \& {Gallo}}{{Gonzalez}
  et~al.}{2017}]{Gonzalez17}
{Gonzalez} A.~G.,  {Wilkins} D.~R.,   {Gallo} L.~C.,  2017, \mn@doi [\mnras]
  {10.1093/mnras/stx2080}, \href
  {https://ui.adsabs.harvard.edu/abs/2017MNRAS.472.1932G} {472, 1932}

\bibitem[\protect\citeauthoryear{{Gonzalez}, {Gallo}, {Kosec}, {Fabian},
  {Alston}, {Berton}  \& {Wilkins}}{{Gonzalez} et~al.}{2020}]{Gonzalez20}
{Gonzalez} A.~G.,  {Gallo} L.~C.,  {Kosec} P.,  {Fabian} A.~C.,  {Alston}
  W.~N.,  {Berton} M.,   {Wilkins} D.~R.,  2020, \mn@doi [\mnras]
  {10.1093/mnras/staa1735}, \href
  {https://ui.adsabs.harvard.edu/abs/2020MNRAS.496.3708G} {496, 3708}

\bibitem[\protect\citeauthoryear{{Goodman} \& {Weare}}{{Goodman} \&
  {Weare}}{2010}]{mcmc}
{Goodman} J.,  {Weare} J.,  2010, \mn@doi [Communications in Applied
  Mathematics and Computational Science] {10.2140/camcos.2010.5.65}, \href
  {https://ui.adsabs.harvard.edu/abs/2010CAMCS...5...65G} {5, 65}

\bibitem[\protect\citeauthoryear{{Grupe}}{{Grupe}}{2004}]{Grupe04b}
{Grupe} D.,  2004, \mn@doi [\aj] {10.1086/382516}, \href
  {https://ui.adsabs.harvard.edu/abs/2004AJ....127.1799G} {127, 1799}

\bibitem[\protect\citeauthoryear{{Grupe}, {Wills}, {Leighly}  \&
  {Meusinger}}{{Grupe} et~al.}{2004}]{Grupe04a}
{Grupe} D.,  {Wills} B.~J.,  {Leighly} K.~M.,   {Meusinger} H.,  2004, \mn@doi
  [\aj] {10.1086/380233}, \href
  {https://ui.adsabs.harvard.edu/abs/2004AJ....127..156G} {127, 156}

\bibitem[\protect\citeauthoryear{{Harrison} et~al.,}{{Harrison}
  et~al.}{2013}]{nustar}
{Harrison} F.~A.,  et~al., 2013, \mn@doi [\apj] {10.1088/0004-637X/770/2/103},
  \href {https://ui.adsabs.harvard.edu/abs/2013ApJ...770..103H} {770, 103}

\bibitem[\protect\citeauthoryear{{Humphrey}, {Liu}  \& {Buote}}{{Humphrey}
  et~al.}{2009}]{Humphrey09}
{Humphrey} P.~J.,  {Liu} W.,   {Buote} D.~A.,  2009, \mn@doi [\apj]
  {10.1088/0004-637X/693/1/822}, \href
  {https://ui.adsabs.harvard.edu/abs/2009ApJ...693..822H} {693, 822}

\bibitem[\protect\citeauthoryear{{Intema}, {Jagannathan}, {Mooley}  \&
  {Frail}}{{Intema} et~al.}{2017}]{TGSS}
{Intema} H.~T.,  {Jagannathan} P.,  {Mooley} K.~P.,   {Frail} D.~A.,  2017,
  \mn@doi [\aap] {10.1051/0004-6361/201628536}, \href
  {https://ui.adsabs.harvard.edu/abs/2017A&A...598A..78I} {598, A78}

\bibitem[\protect\citeauthoryear{{Jiang} et~al.,}{{Jiang}
  et~al.}{2018a}]{Jiang18}
{Jiang} J.,  et~al., 2018a, \mn@doi [\mnras] {10.1093/mnras/sty836}, \href
  {https://ui.adsabs.harvard.edu/abs/2018MNRAS.477.3711J} {477, 3711}

\bibitem[\protect\citeauthoryear{{Jiang}, {Walton}, {Parker}  \&
  {Fabian}}{{Jiang} et~al.}{2018b}]{JiangWKK}
{Jiang} J.,  {Walton} D.~J.,  {Parker} M.~L.,   {Fabian} A.~C.,  2018b, \mn@doi
  [\mnras] {10.1093/mnras/sty2344}, \href
  {https://ui.adsabs.harvard.edu/abs/2018MNRAS.481..639J} {481, 639}

\bibitem[\protect\citeauthoryear{{Jiang}, {Dauser}, {Fabian}, {Alston},
  {Gallo}, {Parker}  \& {Reynolds}}{{Jiang} et~al.}{2022}]{Jiang22}
{Jiang} J.,  {Dauser} T.,  {Fabian} A.~C.,  {Alston} W.~N.,  {Gallo} L.~C.,
  {Parker} M.~L.,   {Reynolds} C.~S.,  2022, \mn@doi [\mnras]
  {10.1093/mnras/stac1144}, \href
  {https://ui.adsabs.harvard.edu/abs/2022MNRAS.tmp.1109J} {}

\bibitem[\protect\citeauthoryear{{Kaastra} \& {Bleeker}}{{Kaastra} \&
  {Bleeker}}{2016}]{optbin}
{Kaastra} J.~S.,  {Bleeker} J.~A.~M.,  2016, \mn@doi [\aap]
  {10.1051/0004-6361/201527395}, \href
  {https://ui.adsabs.harvard.edu/abs/2016A&A...587A.151K} {587, A151}

\bibitem[\protect\citeauthoryear{Kass \& Raftery}{Kass \& Raftery}{1995}]{dic3}
Kass R.~E.,  Raftery A.~E.,  1995, \mn@doi [Journal of the American Statistical
  Association] {10.1080/01621459.1995.10476572}, 90, 773

\bibitem[\protect\citeauthoryear{{Lasota}, {Abramowicz}, {Chen}, {Krolik},
  {Narayan}  \& {Yi}}{{Lasota} et~al.}{1996}]{Lasota}
{Lasota} J.~P.,  {Abramowicz} M.~A.,  {Chen} X.,  {Krolik} J.,  {Narayan} R.,
  {Yi} I.,  1996, \mn@doi [\apj] {10.1086/177137}, \href
  {https://ui.adsabs.harvard.edu/abs/1996ApJ...462..142L} {462, 142}

\bibitem[\protect\citeauthoryear{{Lobban}, {Reeves}, {Porquet}, {Braito},
  {Markowitz}, {Miller}  \& {Turner}}{{Lobban} et~al.}{2010}]{Lobban10}
{Lobban} A.~P.,  {Reeves} J.~N.,  {Porquet} D.,  {Braito} V.,  {Markowitz} A.,
  {Miller} L.,   {Turner} T.~J.,  2010, \mn@doi [\mnras]
  {10.1111/j.1365-2966.2010.17143.x}, \href
  {https://ui.adsabs.harvard.edu/abs/2010MNRAS.408..551L} {408, 551}

\bibitem[\protect\citeauthoryear{{Magdziarz}, {Blaes}, {Zdziarski}, {Johnson}
  \& {Smith}}{{Magdziarz} et~al.}{1998}]{wc1}
{Magdziarz} P.,  {Blaes} O.~M.,  {Zdziarski} A.~A.,  {Johnson} W.~N.,   {Smith}
  D.~A.,  1998, \mn@doi [\mnras] {10.1046/j.1365-8711.1998.02015.x}, \href
  {https://ui.adsabs.harvard.edu/abs/1998MNRAS.301..179M} {301, 179}

\bibitem[\protect\citeauthoryear{{Malizia} et~al.,}{{Malizia}
  et~al.}{2008}]{Malizia08}
{Malizia} A.,  et~al., 2008, \mn@doi [\mnras]
  {10.1111/j.1365-2966.2008.13657.x}, \href
  {https://ui.adsabs.harvard.edu/abs/2008MNRAS.389.1360M} {389, 1360}

\bibitem[\protect\citeauthoryear{{Masetti} et~al.,}{{Masetti}
  et~al.}{2006}]{Masetti06}
{Masetti} N.,  et~al., 2006, \mn@doi [\aap] {10.1051/0004-6361:20066055}, \href
  {https://ui.adsabs.harvard.edu/abs/2006A&A...459...21M} {459, 21}

\bibitem[\protect\citeauthoryear{{Narayan} \& {Yi}}{{Narayan} \&
  {Yi}}{1994}]{ADAF}
{Narayan} R.,  {Yi} I.,  1994, \mn@doi [\apjl] {10.1086/187381}, \href
  {https://ui.adsabs.harvard.edu/abs/1994ApJ...428L..13N} {428, L13}

\bibitem[\protect\citeauthoryear{{Oh} et~al.,}{{Oh} et~al.}{2018}]{Oh18}
{Oh} K.,  et~al., 2018, \mn@doi [\apjs] {10.3847/1538-4365/aaa7fd}, \href
  {https://ui.adsabs.harvard.edu/abs/2018ApJS..235....4O} {235, 4}

\bibitem[\protect\citeauthoryear{{Panessa} et~al.,}{{Panessa}
  et~al.}{2011}]{Panessa11}
{Panessa} F.,  et~al., 2011, \mn@doi [\mnras]
  {10.1111/j.1365-2966.2011.19268.x}, \href
  {https://ui.adsabs.harvard.edu/abs/2011MNRAS.417.2426P} {417, 2426}

\bibitem[\protect\citeauthoryear{{Petrucci}, {Merloni}, {Fabian}, {Haardt}  \&
  {Gallo}}{{Petrucci} et~al.}{2001}]{Petrucci01}
{Petrucci} P.~O.,  {Merloni} A.,  {Fabian} A.,  {Haardt} F.,   {Gallo} E.,
  2001, \mn@doi [\mnras] {10.1046/j.1365-8711.2001.04897.x}, \href
  {https://ui.adsabs.harvard.edu/abs/2001MNRAS.328..501P} {328, 501}

\bibitem[\protect\citeauthoryear{{Petrucci}, {Ursini}, {De Rosa}, {Bianchi},
  {Cappi}, {Matt}, {Dadina}  \& {Malzac}}{{Petrucci} et~al.}{2018}]{Petrucci18}
{Petrucci} P.~O.,  {Ursini} F.,  {De Rosa} A.,  {Bianchi} S.,  {Cappi} M.,
  {Matt} G.,  {Dadina} M.,   {Malzac} J.,  2018, \mn@doi [\aap]
  {10.1051/0004-6361/201731580}, \href
  {https://ui.adsabs.harvard.edu/abs/2018A&A...611A..59P} {611, A59}

\bibitem[\protect\citeauthoryear{{Ptak}, {Yaqoob}, {Mushotzky}, {Serlemitsos}
  \& {Griffiths}}{{Ptak} et~al.}{1998}]{Ptak98}
{Ptak} A.,  {Yaqoob} T.,  {Mushotzky} R.,  {Serlemitsos} P.,   {Griffiths} R.,
  1998, \mn@doi [\apjl] {10.1086/311444}, \href
  {https://ui.adsabs.harvard.edu/abs/1998ApJ...501L..37P} {501, L37}

\bibitem[\protect\citeauthoryear{{Ptak}, {Terashima}, {Ho}  \&
  {Quataert}}{{Ptak} et~al.}{2004}]{Ptak04}
{Ptak} A.,  {Terashima} Y.,  {Ho} L.~C.,   {Quataert} E.,  2004, \mn@doi [\apj]
  {10.1086/382940}, \href
  {https://ui.adsabs.harvard.edu/abs/2004ApJ...606..173P} {606, 173}

\bibitem[\protect\citeauthoryear{{Reynolds}}{{Reynolds}}{2021}]{Reynolds21}
{Reynolds} C.~S.,  2021, \mn@doi [\araa] {10.1146/annurev-astro-112420-035022},
  \href {https://ui.adsabs.harvard.edu/abs/2021ARA&A..59..117R} {59}

\bibitem[\protect\citeauthoryear{{Reynolds}, {Nowak}, {Markoff}, {Tueller},
  {Wilms}  \& {Young}}{{Reynolds} et~al.}{2009}]{Reynolds09}
{Reynolds} C.~S.,  {Nowak} M.~A.,  {Markoff} S.,  {Tueller} J.,  {Wilms} J.,
  {Young} A.~J.,  2009, \mn@doi [\apj] {10.1088/0004-637X/691/2/1159}, \href
  {https://ui.adsabs.harvard.edu/abs/2009ApJ...691.1159R} {691, 1159}

\bibitem[\protect\citeauthoryear{{Ross} \& {Fabian}}{{Ross} \&
  {Fabian}}{2005}]{RossFabian05}
{Ross} R.~R.,  {Fabian} A.~C.,  2005, \mn@doi [\mnras]
  {10.1111/j.1365-2966.2005.08797.x}, \href
  {https://ui.adsabs.harvard.edu/abs/2005MNRAS.358..211R} {358, 211}

\bibitem[\protect\citeauthoryear{{Shafee}, {McClintock}, {Narayan}, {Davis},
  {Li}  \& {Remillard}}{{Shafee} et~al.}{2006}]{Shafee06}
{Shafee} R.,  {McClintock} J.~E.,  {Narayan} R.,  {Davis} S.~W.,  {Li} L.-X.,
  {Remillard} R.~A.,  2006, \mn@doi [\apjl] {10.1086/498938}, \href
  {https://ui.adsabs.harvard.edu/abs/2006ApJ...636L.113S} {636, L113}

\bibitem[\protect\citeauthoryear{{Shakura} \& {Sunyaev}}{{Shakura} \&
  {Sunyaev}}{1973}]{Shakura73}
{Shakura} N.~I.,  {Sunyaev} R.~A.,  1973, \aap, \href
  {https://ui.adsabs.harvard.edu/abs/1973A&A....24..337S} {500, 33}

\bibitem[\protect\citeauthoryear{Spiegelhalter, Best, Carlin  \& Van
  Der~Linde}{Spiegelhalter et~al.}{2002}]{dic}
Spiegelhalter D.~J.,  Best N.~G.,  Carlin B.~P.,   Van Der~Linde A.,  2002,
  \mn@doi [Journal of the Royal Statistical Society: Series B (Statistical
  Methodology)] {https://doi.org/10.1111/1467-9868.00353}, 64, 583

\bibitem[\protect\citeauthoryear{{Thorne}}{{Thorne}}{1974}]{Thorne74}
{Thorne} K.~S.,  1974, \mn@doi [\apj] {10.1086/152991}, \href
  {https://ui.adsabs.harvard.edu/abs/1974ApJ...191..507T} {191, 507}

\bibitem[\protect\citeauthoryear{{Waddell} \& {Gallo}}{{Waddell} \&
  {Gallo}}{2020}]{Waddell20}
{Waddell} S.~G.~H.,  {Gallo} L.~C.,  2020, \mn@doi [\mnras]
  {10.1093/mnras/staa2783}, \href
  {https://ui.adsabs.harvard.edu/abs/2020MNRAS.498.5207W} {498, 5207}

\bibitem[\protect\citeauthoryear{{Waddell} \& {Gallo}}{{Waddell} \&
  {Gallo}}{2022}]{Waddell22}
{Waddell} S.~G.~H.,  {Gallo} L.~C.,  2022, \mn@doi [\mnras]
  {10.1093/mnras/stab3695}, \href
  {https://ui.adsabs.harvard.edu/abs/2022MNRAS.510.4370W} {510, 4370}

\bibitem[\protect\citeauthoryear{{Wayth} et~al.,}{{Wayth} et~al.}{2015}]{GLEAM}
{Wayth} R.~B.,  et~al., 2015, \mn@doi [\pasa] {10.1017/pasa.2015.26}, \href
  {https://ui.adsabs.harvard.edu/abs/2015PASA...32...25W} {32, e025}

\bibitem[\protect\citeauthoryear{{Wilkins} \& {Fabian}}{{Wilkins} \&
  {Fabian}}{2012}]{WF12}
{Wilkins} D.~R.,  {Fabian} A.~C.,  2012, \mn@doi [\mnras]
  {10.1111/j.1365-2966.2012.21308.x}, \href
  {https://ui.adsabs.harvard.edu/abs/2012MNRAS.424.1284W} {424, 1284}

\bibitem[\protect\citeauthoryear{{Wilkins} \& {Gallo}}{{Wilkins} \&
  {Gallo}}{2015a}]{WG15Comp}
{Wilkins} D.~R.,  {Gallo} L.~C.,  2015a, \mn@doi [\mnras]
  {10.1093/mnras/stu2524}, \href
  {https://ui.adsabs.harvard.edu/abs/2015MNRAS.448..703W} {448, 703}

\bibitem[\protect\citeauthoryear{{Wilkins} \& {Gallo}}{{Wilkins} \&
  {Gallo}}{2015b}]{WG15}
{Wilkins} D.~R.,  {Gallo} L.~C.,  2015b, \mn@doi [\mnras]
  {10.1093/mnras/stv162}, \href
  {https://ui.adsabs.harvard.edu/abs/2015MNRAS.449..129W} {449, 129}

\bibitem[\protect\citeauthoryear{{Wilkins}, {Gallo}, {Costantini}, {Brandt}  \&
  {Blandford}}{{Wilkins} et~al.}{2022}]{Wilkins22}
{Wilkins} D.~R.,  {Gallo} L.~C.,  {Costantini} E.,  {Brandt} W.~N.,
  {Blandford} R.~D.,  2022, \mn@doi [\mnras] {10.1093/mnras/stac416}, \href
  {https://ui.adsabs.harvard.edu/abs/2022MNRAS.tmp..417W} {}

\bibitem[\protect\citeauthoryear{{Willingale}, {Starling}, {Beardmore},
  {Tanvir}  \& {O'Brien}}{{Willingale} et~al.}{2013}]{Willingale}
{Willingale} R.,  {Starling} R.~L.~C.,  {Beardmore} A.~P.,  {Tanvir} N.~R.,
  {O'Brien} P.~T.,  2013, \mn@doi [\mnras] {10.1093/mnras/stt175}, \href
  {https://ui.adsabs.harvard.edu/abs/2013MNRAS.431..394W} {431, 394}

\bibitem[\protect\citeauthoryear{{Wilms}, {Allen}  \& {McCray}}{{Wilms}
  et~al.}{2000}]{Wilms}
{Wilms} J.,  {Allen} A.,   {McCray} R.,  2000, \mn@doi [\apj] {10.1086/317016},
  \href {https://ui.adsabs.harvard.edu/abs/2000ApJ...542..914W} {542, 914}

\bibitem[\protect\citeauthoryear{{Younes}, {Ptak}, {Ho}, {Xie}, {Terasima},
  {Yuan}, {Huppenkothen}  \& {Yukita}}{{Younes} et~al.}{2019}]{Younes19}
{Younes} G.,  {Ptak} A.,  {Ho} L.~C.,  {Xie} F.-G.,  {Terasima} Y.,  {Yuan} F.,
   {Huppenkothen} D.,   {Yukita} M.,  2019, \mn@doi [\apj]
  {10.3847/1538-4357/aaf38b}, \href
  {https://ui.adsabs.harvard.edu/abs/2019ApJ...870...73Y} {870, 73}

\bibitem[\protect\citeauthoryear{{Zdziarski}, {Johnson}  \&
  {Magdziarz}}{{Zdziarski} et~al.}{1996}]{nthComp1}
{Zdziarski} A.~A.,  {Johnson} W.~N.,   {Magdziarz} P.,  1996, \mn@doi [\mnras]
  {10.1093/mnras/283.1.193}, \href
  {https://ui.adsabs.harvard.edu/abs/1996MNRAS.283..193Z} {283, 193}

\bibitem[\protect\citeauthoryear{{{\.Z}ycki}, {Done}  \& {Smith}}{{{\.Z}ycki}
  et~al.}{1999}]{nthComp2}
{{\.Z}ycki} P.~T.,  {Done} C.,   {Smith} D.~A.,  1999, \mn@doi [\mnras]
  {10.1046/j.1365-8711.1999.02885.x}, \href
  {https://ui.adsabs.harvard.edu/abs/1999MNRAS.309..561Z} {309, 561}

\makeatother
\end{thebibliography}

\bsp
\label{lastpage}
\end{document}